\documentclass[a4paper,11pt]{article}
\usepackage{jheppub}
\usepackage[dvipsnames]{xcolor}
\usepackage{multirow}
\usepackage{physics}
\usepackage{tabulary}
\usepackage{listings}

\allowdisplaybreaks[4]

\newcommand{\ep}{\ensuremath{\varepsilon}}
\newcommand{\dm}{\ensuremath{\textrm{d}}}

\newcommand{\ijangle}{\ensuremath{\theta}}
\newcommand{\csth}{\ensuremath{\cos \ijangle}}

\newcommand{\emax}{E_{\rm max}}

\newcommand{\Nep}{\ensuremath{\mathcal{N}_{\ep}}}
\newcommand{\intSS}{\ensuremath{\mathcal{S}{\hspace{-5pt}}\mathcal{S}}}
\newcommand{\dInt}[3]{\ensuremath{\Delta_{\texttt{#1}}^{\texttt{#2}}(#3)}}
\newcommand{\tInt}[3]{\ensuremath{\Theta_{\texttt{#1}}}^{\texttt{#2}}(#3)}

\newcommand{\AIz}[1]{\ensuremath{I_{#1}^{(0)}}}
\newcommand{\AIzz}[3]{\ensuremath{I_{#1,#2}^{(0)}\left[#3\right]}}
\newcommand{\AIm}[1]{\ensuremath{I_{#1}^{(1)}}}
\newcommand{\AImz}[2]{\ensuremath{I_{#1,#2}^{(1)}}}

\newcommand{\xa}{\ensuremath{\mathfrak{m}}}
\newcommand{\yb}{\ensuremath{\mathfrak{n}}}

\newcommand{\normA}{\ensuremath{\mathcal{N}_A}}

\newcommand\scalemath[2]{\scalebox{#1}{\mbox{\ensuremath{\displaystyle #2}}}}

\renewcommand{\arraystretch}{1.5}

\def\Siegen{Theoretische Physik 1, Center for Particle Physics Siegen, Universit\"at Siegen, 57068 Siegen, Germany}

\def\KITA{Institute for Theoretical Particle Physics, KIT, Wolfgang-Gaede-Straße 1, 76131, Karlsruhe, Germany}

\preprint{
\begin{flushright}
SI-HEP-2025-08,
TTP25-015,
P3H-25-030
\end{flushright}
}

\title{Integral of the  double-emission eikonal function for  a massive and a massless emitter at an arbitrary angle}
\author[a]{Dennis Horstmann,}
\author[b]{Kirill Melnikov,}
\author[b]{Ming-Ming Long,}
\author[b]{Andrey Pikelner}

\emailAdd{horstmann@physik.uni-siegen.de}
\emailAdd{kirill.melnikov@kit.edu}
\emailAdd{ming-ming.long@kit.edu}
\emailAdd{andrey.pikelner@kit.edu}

\affiliation[a]{\Siegen}
\affiliation[b]{\KITA}

\date{June 2024}

\abstract{ 
We present an analytic calculation of the integrated double-emission eikonal
function of a massive and a massless emitter whose momenta are at an arbitrary
angle to each other. This quantity provides one of the required ingredients for
extending the nested soft-collinear subtraction scheme to processes with massive
final-state particles. To calculate it, we use the standard methodology
involving reverse unitarity and its extension to cases with Heaviside functions,
integration-by-parts technology and reduction to master integrals, and
differential equations. In addition, we also describe a semi-numerical method
based on the subtraction of infra-red and collinear singularities from the
eikonal function, allowing us to extract divergences of the integrated eikonal
function analytically, and to derive a simple integral representation for the
finite remainder.
 }


\allowdisplaybreaks

\begin{document}
\maketitle

\section{Introduction}
Perturbative predictions for various LHC processes have been extended to include
next-to-next-to-leading order QCD corrections in recent years
\cite{Chen:2014gva,Boughezal:2015dra,Caola:2015wna,Chen:2016zka,Campbell:2019gmd,Cacciari:2015jma,Cruz-Martinez:2018rod,Gauld:2021ule,Catani:2022mfv,Chawdhry:2019bji,Chawdhry:2021hkp,Czakon:2020coa,Gauld:2023zlv,Currie:2017eqf,Chen:2022tpk,Badger:2023mgf,Czakon:2021mjy,Czakon:2015owf,Catani:2019hip,Buonocore:2023ljm,Brucherseifer:2014ama,Berger:2016oht,Campbell:2020fhf}.
This progress was made possible by improvements in understanding how to compute
two-loop amplitudes, both analytically \cite{Gehrmann:2011aa, Caola:2014iua,
Abreu:2018zmy, Chawdhry:2020for, Agarwal:2021vdh, Badger:2021ega} and
numerically \cite{ Agarwal:2024jyq, Anastasiou:2022eym,
Aguilera-Verdugo:2020set, Capatti:2020xjc, Kermanschah:2024utt}, and by
continuous refinements of the NNLO subtraction and slicing schemes
~\cite{Frixione:2004is, Gehrmann-DeRidder:2005btv,
Currie:2013vh,Somogyi:2005xz,Somogyi:2006db,Czakon:2010td,Czakon:2011ve,Anastasiou:2003gr,
Caola:2017dug,Catani:2007vq,Grazzini:2017mhc,
Boughezal:2011jf,Gaunt:2015pea,Sborlini:2016hat,Herzog:2018ily,Magnea:2018hab,Bertolotti:2022aih,Capatti:2019ypt,TorresBobadilla:2020ekr}
with the aim of making them efficient and applicable to arbitrary processes.

For the subtraction schemes, one requires integrals of universal soft- and
collinear limits of squared QCD amplitudes, over a phase space of unresolved
partons. Making this statement precise is equivalent to specifying a particular
subtraction scheme, but the need for such ``integrated subtraction terms'' is
scheme-independent. In this paper we consider an integrated double-emission
eikonal function in the context of the nested soft-collinear subtraction scheme
\cite{Caola:2017dug} in a situation where one of the emitters is massive and the
other one is massless, and their momenta are at an arbitrary angle to each
other. We note that a similar computation for two massless emitters whose
momenta are at an arbitrary angle to each other was performed in
Ref.~\cite{Caola:2018pxp}, and for massive back-to-back emitters -- in
Ref.~\cite{Bizon:2020tzr}. To extend the nested soft-collinear subtraction
scheme to processes with arbitrary number of massless and massive particles in
the final state, one needs the integrated double-emission eikonal functions for
the massless-massive and massive-massive emitters, which are at an arbitrary
angle to each other. The goal of this paper is to present the required
calculation for the massless-massive case.

We perform this calculation in two complementary ways. First, we show how to
organize it using reverse unitarity \cite{Anastasiou:2002yz} and its
generalization for integrals with Heaviside functions \cite{Baranowski:2021gxe}.
Reverse unitarity enables the use of multi-loop technology, involving derivation
of linear algebraic identities between different integrals with the help of the
integration-by-parts (IBP) method \cite{Tkachov:1981wb,Chetyrkin:1981qh},
reduction of many integrals contributing to the integrated eikonal to a few
master integrals, and solving differential equations that these integrals
satisfy. Although this methodology allows us to compute the integral of the
double-emission eikonal function analytically, the calculation is quite
challenging. Furthermore, a striking feature of the calculation is that results
for integrals that appear at the intermediate stages are much more complex than
the final result for the integrated double-emission eikonal function. For
example, the system of differential equations for the master integrals that we
require involves elliptic integrals which, however, cancel out in the final
result for the integrated eikonal.

To shed light on this simplicity, we applied a different method to integrate the
eikonal function which relies on constructing subtraction terms for the eikonal
function itself. This approach allows us to extract all singular contributions
from the integrated eikonal and devise a simple representation for the finite
remainder that we then integrate numerically. It is based on the same idea that
was used to compute the NNLO QCD contribution to the $N$-jettiness soft function
in Ref.~\cite{Agarwal:2024gws}, although there are important differences between
the two cases at the practical level.

The paper is organized as follows. In Section~\ref{sec:notation} we introduce
the eikonal functions, as well as other quantities and notations that we need in
the rest of the paper. In Section~\ref{sec:intSnlo} single-emission integrals
and their iterations are discussed. In
Section~\ref{sec:direct-calculation-P-rest} the analytic calculation of the
integrated double-emission eikonal function is reported. In
Section~\ref{sec:calcRestA} an alternative semi-numerical computation of the
same quantity is discussed. We conclude in Section~\ref{sec:concl}. Various
technical details, definitions of useful quantities, as well as instructions on
how to use ancillary files provided with this paper are collected in Appendices.

\section{Notations and definitions}
\label{sec:notation}
We study  a generic partonic process
\begin{equation}
    0 \to h_1(p_1) + \cdots + h_n(p_n) + H_{n+1}(p_{n+1}) + \cdots + H_{N}(p_{N}) + f_1(k_1) + f_2(k_2),
    \label{eq2.1}
\end{equation}
where $h_{i}$ and $H_{i}$ are massless and massive partons, and $f_{1,2}$ are
two massless, potentially unresolved partons which can be either two gluons or a
$q \bar q$ pair. We then consider the double-soft limit, $k_1, k_2 \to 0$, with
all other momenta in Eq.~(\ref{eq2.1}) fixed. In this limit, the amplitude
squared of the process in Eq.~(\ref{eq2.1}) factorizes as follows
\cite{Catani:1999ss}:
\begin{itemize}
\item if $f_{1,2}$ are gluons, 
\begin{equation}
\begin{split}
  \label{eq: double_soft_gg}
 \lim_{k_1,k_2 \to 0} | \mathcal{M}^{gg}(\{p\},k_1,k_2) |^2
  \approx{}
  g_{s,b}^{4}
  \bigg\{ &
            \frac{1}{2} \sum_{i,j,k,l}^{N}
            \mathcal{S}_{ij}(k_1) \mathcal{S}_{kl}(k_2)
            | \mathcal{M}^{\{(ij),(kl)\}}(\{p\}) |^2
            \\
          &
            - C_A \sum_{i,j}^{N} \mathcal{S}_{ij}(k_1,k_2) | \mathcal{M}^{(ij)}(\{p\})|^2
            \bigg\} \ ,
\end{split}
\end{equation}

\item if $f_1 = q$ and $f_2 = \bar q$, 
\begin{align}
  \label{eq: double_soft_qqb}
\lim_{k_1,k_2 \to 0}  | \mathcal{M}^{q\bar{q}}(\{p\},k_1,k_2) |^2
  \approx{}
  g_{s,b}^{4} \ T_R \sum_{i,j}^{N} \mathcal{I}_{ij}(k_1,k_2)
  | \mathcal{M}^{(ij)}(\{p\})|^2  \ .
\end{align}
\end{itemize}

Quantities that appear in the above equations include two Casimir operators of
the $SU(3)$ group, $C_A = 3, T_R = 1/2$, the bare strong coupling constant
$g_{s,b}$, as well as color-correlated matrix elements
\begin{align}
  \label{eq: color_matrix}
  | \mathcal{M}^{\{(ij),(kl)\}}(\{p\})|^2
  &={}
  \langle \mathcal{M}(\{p\}) |
  \{ \boldsymbol{T}_i \cdot \boldsymbol{T}_j ,\boldsymbol{T}_k \cdot \boldsymbol{T}_l\}
  | \mathcal{M}(\{p\}) \rangle \ , \\
  | \mathcal{M}^{\{(ij)\}}(\{p\})|^2
  &={}
  \langle \mathcal{M}(\{p\}) |
  \boldsymbol{T}_i \cdot \boldsymbol{T}_j
  | \mathcal{M}(\{p\}) \rangle \ ,
\end{align}
where $\boldsymbol{T}_i$ are operators of color charges \cite{Catani:1996jh} and
$\{\cdot,\cdot\}$ denotes an anti-commutator. Sums in Eqs. (\ref{eq:
double_soft_gg}, \ref{eq: double_soft_qqb}) run over all pairs of hard
color-charged emitters.

In Eq.~\eqref{eq: double_soft_gg}, the term containing the product of two single-eikonal factors 
\begin{equation}
\label{eq: eikonal_g}
    \mathcal{S}_{ij}(k) = \frac{(p_i \cdot p_j)}{(p_i \cdot k)(p_j \cdot k)} \; ,
\end{equation}
is the \textit{Abelian} contribution. We note that $S_{ij}(k)$ also appears in
the single-emission eikonal contribution relevant for computations at
next-to-leading order.

The \textit{non-Abelian} term, proportional to the color factor $C_A$, is more
complicated. The eikonal function $\mathcal{S}_{ij}(k_1,k_2)$ reads
\begin{align}
  \label{eq: eikonal_gg}
  \mathcal{S}_{ij}(k_1,k_2)
  ={}&
       \mathcal{S}^{0}_{ij}(k_1,k_2) +
       \left[
       m_i^2 \mathcal{S}^{m}_{ij}(k_1,k_2)
       +
       m_j^2 \mathcal{S}^{m}_{ji}(k_1,k_2)
       \right] ,
\end{align}
where quantities that appear in square brackets depend on the masses of the two
emitters, $m_{i,j}$. Although the explicit dependence of the function $S_{ij}$
on emitters' masses is shown in Eq.~(\ref{eq: eikonal_gg}), both functions
$\mathcal{S}^{0}_{ij}(k_1,k_2)$ and $\mathcal{S}^{m}_{ij}(k_1,k_2)$ implicitly
depend on them through their momenta $p_{i,j}$, $p_{i,j}^2 = m_{i,j}^2$.

The first term in
Eq.~\eqref{eq: eikonal_gg}, $\mathcal{S}^{0}_{ij}(k_1,k_2)$ is the same for massless and massive emitters
\cite{Catani:1999ss}. It reads
\begin{equation}
  \begin{split}
    \label{eq: double_eikonal_s0}
    \mathcal{S}^0_{ij}(k_1,k_2) 
    ={}&   \frac{(1-\ep)}{(k_1\cdot k_2)^2} \frac{\left[(p_i\cdot k_1)(p_j\cdot k_2) + i\leftrightarrow j \right]}{(p_i\cdot k_{12})(p_j\cdot k_{12})}  \\
       & - \frac{(p_i\cdot p_j)^2}{2(p_i\cdot k_1)(p_j\cdot k_2)(p_i\cdot k_2)(p_j\cdot k_1)} \bigg[ 2 - \frac{\left[(p_i\cdot k_1)(p_j\cdot k_2) + i\leftrightarrow j \right]}{(p_i\cdot k_{12})(p_j\cdot k_{12})} \bigg] \\
       & + \frac{(p_i\cdot p_j)}{2(k_1\cdot k_2) } \bigg[ \frac{2}{(p_i\cdot k_1)(p_j\cdot k_2)} + \frac{2}{(p_j\cdot k_1)(p_i\cdot k_2)} - \frac1{(p_i\cdot k_{12})(p_j\cdot k_{12})}  \\
       & ~~ \times \left( 4 + \frac{\left[(p_i\cdot k_1)(p_j\cdot k_2) + i\leftrightarrow j \right]^2}{(p_i\cdot k_1)(p_j\cdot k_2)(p_i\cdot k_2)(p_j\cdot k_1)} \right) \bigg] \ ,
  \end{split}
\end{equation}
where we have used the abbreviation $k_{12} = k_1+k_2$. The other two
contributions in Eq.~\eqref{eq: eikonal_gg} are only relevant for massive
emitters. The function $\mathcal{S}^{m}_{ij}(k_1,k_2)$ is given
by~\cite{Catani:2019nqv}\footnote{ We note that a somewhat different expression
for $S_{ij}^{m}$ is given in Ref.~\cite{Czakon:2011ve}. However, both
expressions give the same result after summing over $i,j$ in Eqs.~(\ref{eq:
double_soft_gg}, \ref{eq: double_soft_qqb}) thanks to colour conservation. }
\begin{equation}
  \begin{split}
    \label{eq: double_eikonal_sm}
    \mathcal{S}^{m}_{ij}(k_1,k_2) 
    ={}& \frac{(p_i\cdot p_j) (p_j \cdot k_{12})}{2(p_i\cdot k_1)(p_j\cdot k_2)(p_i\cdot k_2)(p_j\cdot k_1)(p_i \cdot k_{12})} \\
       & - \frac1{2(k_1\cdot k_2)(p_i\cdot k_{12})(p_j\cdot k_{12})} \left( \frac{(p_j\cdot k_1)^2}{(p_i\cdot k_1)(p_j\cdot k_2)} + \frac{(p_j\cdot k_2)^2}{(p_i\cdot k_2)(p_j\cdot k_1)} \right).
  \end{split}
\end{equation}

In the quark-antiquark  case, the eikonal function $\mathcal{I}_{ij}(k_1,k_2)$ reads
\begin{align}
  \label{eq: eikonal_qqb}
  \mathcal{I}_{ij}(k_1,k_2) = \frac{\left[ \left(p_i \cdot k_1 \right) \left(p_j \cdot k_2 \right) + i\leftrightarrow j \right] -\left(p_i \cdot p_j \right)\left(k_1 \cdot k_2 \right) }{\left(k_1 \cdot k_2 \right)^2\left(p_i \cdot k_{12} \right) \left( p_j \cdot k_{12} \right)} \ ,
\end{align}
and there is no difference between massive and massless emitters. 

It turns out to be  convenient to make use of color conservation
\begin{equation}
  \sum \limits_{i=1}^{N} \boldsymbol{T}_i |{\cal M}(\{p\}) \rangle  = 0,
\end{equation}
and the symmetry of functions $S_{ij} = S_{ji}$ and $I_{ij} = I_{ji}$ to write
\begin{align}
  \sum_{i,j}^{N} \mathcal{S}_{ij}(k_1,k_2) | \mathcal{M}^{(ij)}(\{p\})|^2 &= \sum_{i<j}^{N} \widetilde{\mathcal{S}}_{ij}(k_1,k_2) | \mathcal{M}^{(ij)}(\{p\})|^2, \\
  \sum_{i,j}^{N} \mathcal{I}_{ij}(k_1,k_2) | \mathcal{M}^{(ij)}(\{p\})|^2 &= \sum_{i<j}^{N} \widetilde{\mathcal{I}}_{ij}(k_1,k_2) | \mathcal{M}^{(ij)}(\{p\})|^2,
\end{align}
where 
\begin{align}
  \label{eq:tildedEikDef}
  \widetilde{\mathcal{S}}_{ij} & = 2\mathcal{S}_{ij}  - \mathcal{S}_{ii} - \mathcal{S}_{jj},\\
  \widetilde{\mathcal{I}}_{ij} & = 2\mathcal{I}_{ij}  - \mathcal{I}_{ii} - \mathcal{I}_{jj}.
\end{align}
Functions $\tilde S_{ij}, \tilde I_{ij}$ are more suitable for the analysis
presented below, especially in Section~\ref{sec:calcRestA}, because they only
have physical singularities for each pair $i,j$.

To compute the required double-soft contribution, we have to integrate the
corresponding eikonal functions $\tilde S_{ij}$ and $\tilde I_{ij}$ over the
phase space of two unresolved partons with momenta $k_{1,2}$. Working within the
nested soft-collinear subtraction scheme, we have to fix the reference frame and
restrict energies of unresolved partons by introducing an upper cut-off $\emax$.
Furthermore, energies of unresolved partons must be ordered. We call the parton
with the larger (smaller) energy $\xa(\yb)$, and refer to their momenta as
$k_{\xa,\yb}$, instead of $k_{1,2}$, which describe momenta without energy
ordering.

We define the  required double-emission phase-space integrals as
\cite{Caola:2017dug}
\begin{align}
  \label{eq:ssintDef1}
  \intSS\left[ \mathcal{S}_{ij} \mathcal{S}_{kl} \right] &= \int [\dm k_{\xa}] [\dm k_{\yb}]
  \theta\left(\emax - k_{\xa}^0  \right)
  \theta\left(k_{\xa}^0 - k_{\yb}^0  \right)
  \mathcal{S}_{ij}(k_{\xa}) \mathcal{S}_{kl}(k_{\yb}), \\ \label{eq:ssintDef2}
  \intSS\left[ \Xi_{ij} \right] &= \int [\dm k_{\xa}] [\dm k_{\yb}]
  \theta\left(\emax - k_{\xa}^0  \right)
  \theta\left(k_{\xa}^0 - k_{\yb}^0  \right)
  \Xi_{ij}\left(k_{\xa},k_{\yb}\right), 
\end{align}
where the eikonal function $\Xi_{ij}$ is either $\tilde{\mathcal{S}}_{ij}$ (for
$gg$ emission) or $\tilde{\mathcal{I}}_{ij}$ (for $q \bar q$ emission),
\begin{equation}
    [\dm k] = \frac{\dm^{d-1}{k}}{2 k^0 (2\pi)^{d-1} },
\end{equation}
is the phase-space element and $d$ is the space-time dimension
$d=4-2\ep$.\footnote{We use dimensional regularization to regulate soft and
collinear divergences throughout this paper.}

The dependence of soft integrals on $\emax$ can be made manifest. We use
Eq.~\eqref{eq:ssintDef2} to demonstrate it. We use the integral representation
for the first theta-function
\begin{equation}
  \label{eq:thetaInt}
  \theta(b-a) = \int\limits_0^1 \dm z \ \delta(z b -a) \ b,
\end{equation}
and obtain
\begin{equation}
  \label{eq:ssintNoTh}
  \intSS\left[ \Xi_{ij} \right] = \int\limits_0^1 \frac{\dm z}{z} \int [\dm k_{\xa}] [\dm k_{\yb}]
  \delta\left(1 - \frac{k_{\xa}^0}{z \emax}  \right)
  \theta\left(k_{\xa}^0 - k_{\yb}^0  \right)
  \Xi_{ij}\left(k_{\xa},k_{\yb}\right).
\end{equation}
Since the eikonal functions are homogeneous, i.e. $\Xi_{ij}(\lambda k_{\xa},
\lambda k_{\yb}) = \lambda^{-4} \Xi_{ij}(\{k_{\xa}, k_{\yb})$, and since the
integration measure satisfies $[\dm(\lambda k_i)] \to \lambda^{d-2} [\dm k_i]$,
we rescale both momenta $k_{\xa,\yb}$ with $\lambda = z \emax$ and find
\begin{equation}
  \begin{split}
    \label{eq:ssintNoThKtoL}
    \intSS\left[ \Xi_{ij} \right] & = \int\limits_0^1 \frac{\dm z}{z} \frac{(z \emax)^{2(d-2)}}{\left( z \emax \right)^4}
                                    \int [\dm l_{\xa}] [\dm l_{\yb}]
                                    \delta\left(1 - l_{\xa}\cdot P\right)
                                    \theta\left(l_{\xa}\cdot P - l_{\yb}\cdot P  \right)
                                    \Xi_{ij}\left(l_{\xa},l_{\yb}\right)
    \\
                                  & = - \frac{1}{4\ep \emax^{4\ep}}
                                    \int [\dm l_{\xa}] [\dm l_{\yb}]
                                    \delta\left(1 - l_{\xa}\cdot P\right)
                                    \theta\left(l_{\xa}\cdot P - l_{\yb}\cdot P  \right)
                                    \Xi_{ij}\left(l_{\xa},l_{\yb}\right).
  \end{split}
\end{equation}
We note that an auxiliary four-vector $P=(1,\vec{0})$ was introduced in
Eq.~(\ref{eq:ssintNoThKtoL}), to ensure that energies of unresolved partons are
constrained. A similar formula is easily obtained for the integral of the
product of two single eikonal functions $\intSS[S_{ij} S_{kl}]$.

Our goal is to compute the required soft integrals for a massive ($i$) and a
massless $(j)$ emitter. We label their momenta as $p_{i,j}$. If $m$ is the mass
parameter, then $p_i^2 = m^2$, and $p_j^2 = 0$. In the ``laboratory frame''
where $P=(1,\vec{0})$, we employ the following momenta parametrization
\begin{equation}
    p_i = E_i (1, \beta \vec{n}_i), \qquad
    p_j = E_j (1, \vec{n}_j),
\end{equation}
where the three-vectors $\vec n_{i,j}$ satisfy $\vec{n}_{i,j}^2=1$, and $\beta$
is the velocity of the massive emitter
\begin{equation}
  \beta = \sqrt{1-\frac{m^2}{E_i^2}}.
\end{equation}

In principle,  the  integrated eikonal function can  depend on the following scalar products
\begin{gather}\label{eq:spDef}
      p_i^2 = m^2,  \quad P^2 = 1,\quad
      p_i\cdot p_j = E_i E_j (1-\beta\csth), \quad p_{i,j} \cdot P = E_{i,j}, 
\end{gather}
where $\ijangle$ is the angle between $\vec n_i$ and $\vec n_j$, i.e. $\vec n_i
\cdot \vec n_j = \csth$. Since the eikonal functions are homogeneous in
$p_{i,j}$, the dependencies on $E_{i,j}$ cancel out, so that in practice
integrals $\intSS$ are functions of $\beta$, $\csth$, and the energy cut-off
$\emax$. The dependence on $\emax$ has already been found, see
Eq.~(\ref{eq:ssintNoThKtoL}); thus, integrated eikonal functions have
non-trivial functional dependence on $\beta$ and $\cos \theta$ only.

\section{Single emission integrals and their iterations}
\label{sec:intSnlo}

We start by   considering the product of two single-soft eikonal functions,
\begin{equation}
    \intSS\left[ \mathcal{S}_{ij} \mathcal{S}_{kl}\right]
    =
    - \frac{1}{4\ep \emax^{4\ep}}
  \int [\dm l_{\xa}] [\dm l_{\yb}]
  \delta\left(1 - l_{\xa}\cdot P\right)
  \theta\left(l_{\xa}\cdot P - l_{\yb}\cdot P  \right)
  \mathcal{S}_{ij}(l_{\xa}) \mathcal{S}_{kl}(l_{\yb}).
\end{equation}
We  resolve  the constraint imposed by the $\theta$-function by writing 
\begin{equation}
\label{eq: para_energy}
    l_{\yb} \cdot P = t \; \, l_{\xa} \cdot P, \quad 0 < t <1.
\end{equation}
Integration over $t$ and the energy component $P \cdot l_\xa$ of parton $\xa$ is
elementary. We obtain
\begin{equation}
\label{eq: ssintRes1}
    \intSS\left[ \mathcal{S}_{ij}(k_{\xa}) \mathcal{S}_{kl}(k_{\yb}) \right]
    =
    \frac{\mathcal{G}_{ij} \; \mathcal{G}_{kl}}{8\ep^2 \emax^{4\ep}},
\end{equation}
where $\mathcal{G}_{ij}, \mathcal{G}_{kl}$ are angular integrals which we write
in the following way ($\mathcal{G}_{kl}$ is the same function up to $p_i \to
p_k$, $p_j \to p_l$ reassignment),
\begin{equation}
\mathcal{G}_{ij} = \int \frac{\dd{\Omega_{\xa}^{(d-1)}}}{2\,(2\pi)^{d-1}} \frac{1-\beta \cos \theta }{(1 - \beta \vec n_i \cdot \vec n_\xa)
(1 - \vec n_j  \cdot \vec n_{\xa})  },
\label{eq3.4}
\end{equation}
where integration over directions of the $(d-1)$-dimensional unit vector $\vec
n_{\xa}$ is to be performed. We note that this integral can be expressed through
an Appell function to all orders in $\ep$ \cite{Somogyi:2011ir}, see
Appendix~\ref{sec:angIntsDef}. However, below we explain how to compute it using
multi-loop methodology including integration-by-parts, reduction to master
integrals and differential equations.

To compute the integral in Eq.~(\ref{eq3.4}), we employ reverse unitarity
\cite{Anastasiou:2002yz} and other standard methods for multi-loop computations,
such as integration-by-parts \cite{Tkachov:1981wb,Chetyrkin:1981qh} and
differential equations \cite{Kotikov:1990kg,Gehrmann:1999as,Henn:2013pwa}. To
this end, we consider a family of integrals~\footnote{As we will see in
Section~\ref{sec:calcRestA}, we need to consider the case where the power of the
``propagators'' becomes $\ep$-dependent. In general, such integrals can be
expressed through Appell functions, see Appendix~\ref{sec:angIntsDef}. In this
section, we focus on integer powers of propagators only.}
\begin{equation}
    I(a_1,a_2) \equiv \int \frac{\dd{\Omega_{\xa}^{(d-1)}}}{2\,(2\pi)^{d-1}} \frac{(1 - \beta \csth )}{(1 -  \beta\vec{n}_i \cdot \vec{n}_{\xa})^{a_1} ( 1 - \vec{n}_j \cdot \vec{n}_{\xa})^{a_2}} \; , \quad a_i \in \mathbb{Z},
\end{equation}
that generalize $\mathcal{G}_{ij} = I(1,1)$. We would like to derive the
differential equation for $I(1,1)$ with respect to $\beta$. To do this, we write
the solid angle element as
\begin{equation}
    \dd{\Omega_{\xa}^{(d-1)}} = 2\,\dd[d]{l_{\xa}} \delta^+(l_{\xa}^2) \, \delta(1-l_{\xa} \cdot P) \; ,
\end{equation}
and proceed with using reverse unitarity to rewrite $\delta$-functions as
``propagators''. This leads to the differential equation
\begin{equation}
\label{eq3.9}
    \pdv{I(1,1)}{\beta} = \frac{1}{ 1 - \beta \csth} \left[\frac{1 - 2\ep}{(1-\beta^2)\beta} \;  I(0,0) + \frac{2\ep \beta }{1-\beta^2}\; I(1,0) + \frac{2\ep}{\beta}\; I(1,1) \right]\; .
\end{equation}

Calculating integrals $I(0,0)$ and $I(1,0)$, and choosing the point $\beta = 0$
to compute the boundary condition,\footnote{We note that at $\beta = 0$, the
dependence of the integral on $\csth$ disappears.} we calculate $I(1,1)$ as a
series expansion in $\ep$. We find
\begin{equation}
\label{eq: mi_nlo_11}
    I(1,1) = \frac{(1-2\ep)}{\ep}\Nep  
    \sum_{n=0} I_{11}^{(n)} \ep^n,
\end{equation}
where the normalization factor is
\begin{equation}
    \Nep = \frac{\pi^{\ep}}{4\pi^2} \frac{\Gamma(1-\ep)}{\Gamma(2-2\ep)},
\end{equation}
and the first three coefficients in Eq.~\eqref{eq: mi_nlo_11} read
\begin{equation}
    \begin{split}
        I_{11}^{(0)} &= -\frac{1}{2}, \\
        I_{11}^{(1)} &= \log \left( \frac{1 - \beta \csth }{\sqrt{1 - \beta^2}} \right), \\
        I_{11}^{(2)} &= - \left[ \frac{1}{4} \log^2 \left( \frac{1-\beta}{1+\beta} \right) + \log \left( \frac{1-\beta \csth}{1+\beta} \right) \log \left( \frac{1-\beta \csth}{1-\beta} \right) \right. \\ 
     & \qquad\qquad + \left. \vphantom{\frac{1}{4}} \text{Li}_2 \left( 1 - \frac{1-\beta \csth}{ 1 + \beta } \right) + \text{Li}_2 \left( 1 - \frac{1-\beta \csth}{ 1 - \beta } \right) \right].
    \end{split}
\end{equation}
We note that they agree with the results reported in
Refs.\,\cite{Bronnum-Hansen:2022tmr, Alioli:2010xd}. With these expressions, it
is straightforward to compute the soft integral in Eq.~(\ref{eq: ssintRes1}).
Judging from Eq.~(\ref{eq: ssintRes1}), it may appear that terms up to
$I_{11}^{(4)}$ will be required to obtain the finite contribution for the
integrated double-soft subtraction term. This, however, may not be necessary
since $\intSS[S_{ij} S_{kl}]$ will get combined with the iterated Catani's
operator $\boldsymbol{I}_1$ \cite{Catani:1996jh,Catani:1998bh} which should
ensure cancellation of $1/\ep$ singularities in the iterated structure, reducing
the depth of the $\ep$-expansion in Eq.~(\ref{eq: mi_nlo_11}) required to obtain
${\cal O}(\ep^0)$ contributions to physical quantities. Nevertheless, for
completeness, we also include higher order $\ep$-expansion terms in
Eq.~(\ref{eq: mi_nlo_11}) in the ancillary files provided with this paper.

\section{ 
Integrating  double-emission eikonal function in the laboratory frame}
\label{sec:direct-calculation-P-rest}

We continue with the discussion of the non-Abelian contribution to the
integrated double-emission eikonal function. Compared to the iterated
single-emission case discussed in Section~\ref{sec:intSnlo}, its calculation is
much more involved. For this reason, it is important to design an efficient
toolchain for calculating integrals appearing in Eq.~\eqref{eq:ssintNoThKtoL}.
The plan is to map these integrals on a small set of master integrals, and use
differential equations to calculate them. We note that a similar approach was
successfully used in calculations with two massless emitters at an arbitrary
angle \cite{Caola:2018pxp}, and with two massive emitters in the back-to-back
limit \cite{Bizon:2020tzr}. However, we will see that it remains challenging to
utilize it in the massive-massless case.

\subsection{IBP reduction and master integrals}
\label{sec:IBPandMI}

The first step in making the computation of the integrated double-emission
eikonal function defined in Eq.~\eqref{eq:ssintNoThKtoL} systematic is to
classify contributing integrals. Since eikonal functions contain linear
propagators, we perform partial fractioning and express all integrals through
the following ones
\begin{equation}
  \label{eq:tIntDef}
  \tInt{abc}{def}{g} = \int
  \frac{[\dm l_{\xa}] [\dm l_{\yb}]\delta\left(1 - l_{\xa}\cdot P\right)
    \theta\left(l_{\xa}\cdot P - l_{\yb}\cdot P  \right)}{
    (l_{\xa}\cdot p_i)^a(l_{\yb}\cdot p_i)^b(l_{\xa \yb}\cdot p_i)^c
    (l_{\xa}\cdot p_j)^d(l_{\yb}\cdot p_j)^e(l_{\xa \yb}\cdot p_j)^f
    (l_{\xa}\cdot l_{\yb})^g
  },
\end{equation}
where $l_{\xa \yb} = l_\xa + l_\yb$. The corresponding integrands contain
quadratic and linear propagators, as well as constraints provided by delta- and
theta-functions. We note that integrals without theta-functions, can be
expressed through master integrals using publicly available codes
\cite{Lee:2012cn,Klappert:2020nbg}. However, if integrands do contain
theta-functions, a generalization of this approach is required.

There are several ways to achieve this. For example, one can write a
theta-function as an integral of a delta-function, c.f. Eq.~(\ref{eq:thetaInt}),
and employ generalized unitarity to deal with integrals that only contain
delta-function constraints afterwards. The drawback of this approach is that the
integrals that we need to calculate become functions of three, rather than two
variables. Although the reduction to master integrals and the derivation of
differential equations are possible with this approach, we find that it leads to
over-complicated intermediate results.

Thankfully, we do not need to follow this approach because there exists an
alternative method for dealing with real-emission integrals that contain
Heaviside functions. It was developed in Ref.~\cite{Baranowski:2021gxe}, and
successfully applied in the calculation of N$3$LO zero-jettiness soft function
\cite{Baranowski:2022khd,Baranowski:2024ene,Baranowski:2024vxg,Baranowski:2024ysi}.

This method is based on a simple observation that action of the differential
operator on the integrand -- a crucial step in establishing the IBP methodology
\cite{Tkachov:1981wb,Chetyrkin:1981qh} -- can be generalized to include
integrands with theta-functions. The key to this generalization is the following
equation
\begin{equation}
  \label{eq:dTheta}
  \frac{\partial}{\partial s}\theta(f(s)) = \delta(f(s)) \frac{\partial}{\partial s} f(s),
\end{equation}
where $s$ is a scalar product constructed from internal and external momenta,
and a typical function $f(s)$ is a linear combination of scalar products between
various four-momenta.

There are two important consequences of Eq.~(\ref{eq:dTheta}). First, each
application of Eq.~\eqref{eq:dTheta} replaces a theta-function with a
delta-function, and it is impossible to express such integrals through
$\Theta$-integrals defined in Eq.~\eqref{eq:tIntDef}. This makes the required
reduction \emph{inhomogeneous}, and we need to extend the set of considered
integrals by including integrals with two delta-functions
\begin{equation}
  \label{eq:dIntDef}
  \dInt{abc}{def}{g} = \int
  \frac{[\dm l_{\xa}] [\dm l_{\yb}]\delta\left(1 - l_{\xa}\cdot P\right)
    \delta\left(1 - l_{\yb}\cdot P  \right)}{
    (l_{\xa}\cdot p_i)^a(l_{\yb}\cdot p_i)^b(l_{\xa \yb}\cdot p_i)^c
    (l_{\xa}\cdot p_j)^d(l_{\yb}\cdot p_j)^e(l_{\xa \yb}\cdot p_j)^f
    (l_{\xa}\cdot l_{\yb})^g
  }.
\end{equation}
 Since such integrals do not contain Heaviside functions, they can be dealt with
using any public code for IBP reduction.
 
Second, once a new delta-function appears in the integrand and is mapped on a
new propagator as required by reverse unitarity, it is not guaranteed that this
propagator is linearly-independent of the ones that are already present in an
integral. To get rid of linearly-dependent propagators, we perform the partial
fraction decomposition, each time Eq.~\eqref{eq:dTheta} is applied.

Although these additional reduction steps are relatively simple when taken
separately, they require significant amount of work to be implemented in a
working code. Luckily, many of these steps were designed and implemented for the
calculation of the N3LO zero-jettiness soft function described in
Ref.~\cite{Baranowski:2024ene}. Using the notion introduced in that paper to
describe the present calculation, we need to consider two reduction levels.
First, we express the original integrals from the eikonal function through the
minimal set of $\Theta$-integrals defined in Eq.~\eqref{eq:tIntDef}, and then
proceed with the level-zero reduction of $\Delta$-integrals shown in
Eq.~\eqref{eq:dIntDef}, which is done with \texttt{Kira}
\cite{Klappert:2020nbg}. After the reduction, we find that we require $52$
master integrals including $37$ of the $\Delta$ type and $15$ of the $\Theta$
type. The full list of master integrals can be found in Appendix~\ref{sec:MIs}.

We note that the application of IBP identities has not resulted in a significant
reduction in the number of integrals that need to be computed. Nevertheless, the
reduction is very useful because it allows us to control the complexity of
integrals that we choose to calculate. Furthermore, an opportunity to express
any integral in Eqs.~\eqref{eq:tIntDef} and \eqref{eq:dIntDef} through master
integrals is crucial for deriving differential equations that these integrals
satisfy. Indeed, after differentiating an integrand of a particular integral
with respect to one of the variables $x = \beta$ or $y=\csth$, the obtained
integrals can be expressed through master integrals again. This leads to
differential equations
\begin{equation}
  \label{eq:DEmatDEf}
  \partial_x \vec{J}(x,y) = M_x \vec{J}(x,y), \quad
  \partial_y \vec{J}(x,y) = M_y \vec{J}(x,y), 
\end{equation}
where $\vec J(x,y)$ is a vector of all master integrals that need to be computed
and $M_{x,y}$ are matrices that satisfy the integrability condition
\begin{equation}
\quad
  \partial_xM_y - \partial_y M_x = \left[M_x, M_y \right].
\end{equation}
Furthermore, linear IBP relations allow construction of the recurrence relations
with respect to space-time dimension $d$. They read
\begin{equation}
  \label{eq:drr}
  \vec{J}(d) = L(d)\vec{J}(d+2),
\end{equation}
thereby connecting integrals computed in $d$ and $d+2$ dimensions.
Eq.~(\ref{eq:drr}) is quite useful for numerical checks, since real-emission
integrals that we are interested in become convergent for sufficiently large
$d$, and can be calculated by a straightforward integration, see
Appendix~\ref{sec:directInt} for details.

\subsection{Solution of the differential equations}
\label{sec:deSol}

In this section we explain how the differential equations in
Eq.~(\ref{eq:DEmatDEf}) are solved. Following common practices, we aim at
constructing solutions in terms of iterated integrals. The best way to do this
is to bring the system into a global $\ep$-form \cite{Henn:2013pwa}, by changing
the basis of integrals if possible. The next-to-best option is to find a basis
where matrices $M_{x,y}$ are block-triangular, which means that each matrix
block is a lower-triangular matrix such that all of its diagonal entries are
proportional to $\ep$. For such matrices $M_{x,y}$, integrals $\vec J$ at a
particular order in the $\ep$-expansion depend on their lower-expansion orders,
and simpler integrals.

Following methods described in Ref.~\cite{Adams:2017tga} it is possible to find
a new set of master integrals such that all \emph{but two} diagonal blocks of
the system of differential equations are in the required form. We note that the
transformation from the old to the new basis involves several complicated square
roots.

\begin{figure}
  \centering
  \begin{tabular}{cc}
    \includegraphics[width=0.4\textwidth]{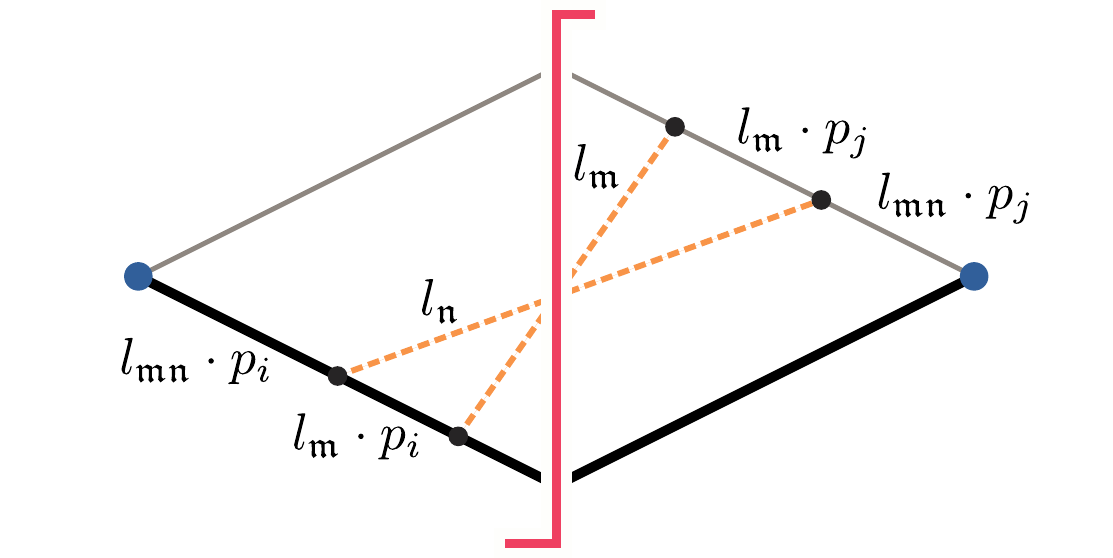}
    & \includegraphics[width=0.4\textwidth]{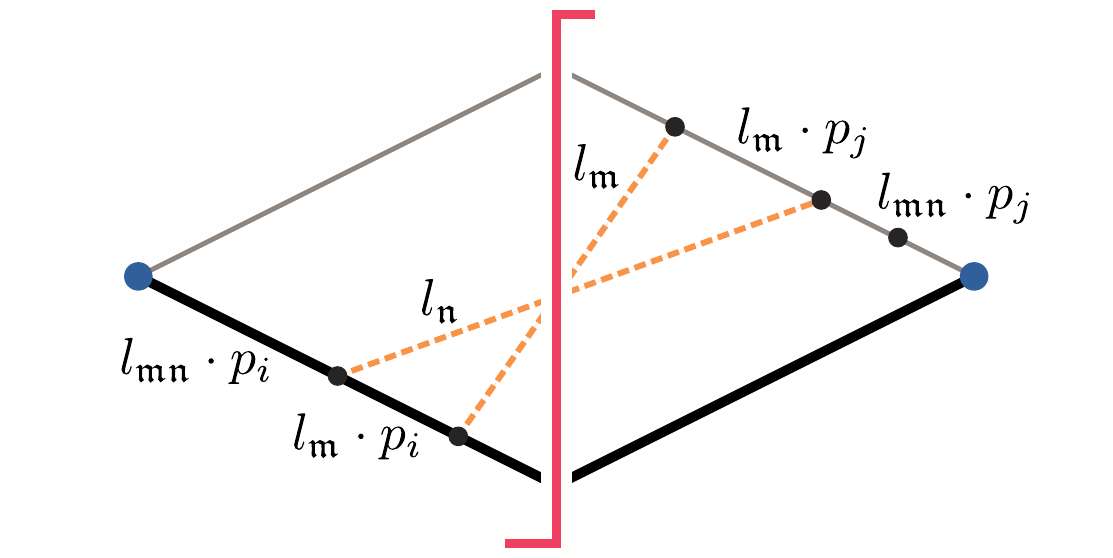}\\
    $J_{30}$ & $J_{31}$
  \end{tabular}
  \caption{Integrals from the sector, which is not first-order factorizable.
    Bold (light) solid lines are massive (massless) eikonal propagators; vertical
    line represent a cut. Particles whose propagators are cut are considered
    on-shell. }
  \label{fig:elInts}
\end{figure}

Finding integrals that appear in the remaining two sectors is one of the major
difficulties of this calculation, and we explain below how this challenge is
addressed. The two irreducible blocks contain four integrals, that we refer to
as $J_{30,31}$ and $J_{34,35}$; they can be found in Appendix~\ref{sec:MIs}. The
homogeneous parts of the matrices $M_{x,y}$ for these blocks are identical;
hence, we focus here on the first pair of integrals shown in
Fig.~\ref{fig:elInts}.

For these integrals the second order differential operator (the Picard-Fuchs
operator) constructed from the homogeneous part of the differential equation
that $J_{30,31}$ satisfy, cannot be factorized into two first order differential
operators. This implies that the solution of the system might involve elliptic
integrals \cite{Adams:2017tga,Adams:2018yfj}. It is useful to consider
differential equations in a single variable $x$, and write them as follows
\begin{equation}
  \label{eq:elDE}
  \partial_x
  \begin{pmatrix}
    J_{30}\\
    J_{31}
  \end{pmatrix} =
  \sum\limits_{i=0}^{2} \ep^i A_i
  \begin{pmatrix}
    J_{30}\\
    J_{31}
  \end{pmatrix}
  + B \vec{J}_s,
\end{equation}
where $A_i$ are $\ep$-independent matrices of the diagonal block and the matrix
$B$ describes the inhomogeneous part of the differential equation. Integrals
$\vec J_s$ are supposed to be known at this point.

Our goal is to construct a transformation to a new set of integrals, satisfying
the system of equations shown in Eq.~\eqref{eq:elDE}, but without the
$\ep$-independent contribution. To accomplish this, we focus on the homogeneous
part of Eq.~(\ref{eq:elDE}) and start by truncating it at ${\cal O}(\ep^0)$. The
matrix $A_0$ reads
\begin{equation}
  \label{eq:elMat}
  A_0=
  \begin{pmatrix}
    \frac{1}{2(1-x)} - \frac{1}{x} - \frac{1}{2 (1 + x)},
    &
      -\frac{2}{x} - \frac{1 - y}{1 - x} + \frac{1 + y}{1 + x}
    \\
    \frac{1}{4(1 - x)(1 - y)} - \frac{1}{4(1 + x)(1 + y)} - \frac{y}{2(1 - y^2)(1 - x y)},
    &
      - \frac{1}{2(1 - x)} + \frac{1}{2(1 + x)} - \frac{1}{x - y} + \frac{2y}{1 - x y}
  \end{pmatrix}.
\end{equation}
We use it to turn a  system of first-order differential equations into a single second order differential equation. It reads 
\begin{equation}
  \label{eq:elOneDE}
\partial_x^2 h + p_1(x,y) \partial_x h + p_0(x,y) h = 0,
\end{equation}
where $h = h(x,y)$ is a ``homogeneous'' version of the integral $J_{30}$, and
the coefficients $p_{1,0}(x,y)$ are given by
\begin{equation}
  \begin{split}
    \label{eq:oneDEcoeff}
    p_1(x,y) & = \frac{y}{x y-1}+\frac{1}{x-y}+\frac{1}{x-1}+\frac{2}{x}+\frac{1}{x+1},\\
    p_0(x,y) & = \frac{y^2}{x y-1}-\frac{y}{x} +\frac{1}{x (x-y)}
               +\frac{1}{2 (x-1) (x-y)}+ \frac{1}{2 (x+1) (x-y)}\\
             & +\frac{1}{2 (x-1) (x y-1)}-\frac{1}{2 (x+1) (x y-1)}+\frac{3}{2(x-1)}-\frac{3}{2 (x+1)}.
  \end{split}
\end{equation}
To simplify Eq.~(\ref{eq:elOneDE}), we define a new  function  
\begin{equation}
  g(x,y) = x(1-xy) h(x,y),
\end{equation}
and re-write the differential equation  through a variable  $\lambda$ defined as 
\begin{equation}
  \label{eq:lamDef}
  \lambda = \frac{(1-x^2)(1-y^2)}{(1-x y)^2}.
\end{equation}
The transformed equation reads
\begin{equation}
  \label{eq:gKeq}
4 \lambda(1-\lambda)\; \partial_\lambda^2
  g
    + 4(1-2\lambda) \; \partial_\lambda g 
  - g  = 0.
\end{equation}
This equation is the second order differential equation that defines the
complete elliptic integral of the first kind. It has two independent solutions
$g(\lambda) = K(\lambda)$ and $g(\lambda) = K(1-\lambda)$ where $K(\lambda)$ is
defined as follows
\begin{equation}
  \label{eq:Kdef}
  K(\lambda) = \int\limits_0^1 \frac{\dm t}{\sqrt{1-t^2}\sqrt{1-\lambda t^2}}.
\end{equation}
Interestingly, the parameter $\lambda$ can be written as a sine squared of an
angle $\sigma$ which appears to be an angle between directions of vectors $\vec
P$ and $\vec p_j$ \emph{in the rest frame of the massive parton $p_i$}. We note
that we use the rest frame of the parton $i$ to set up a semi-numerical
computation of the integrated double-emission eikonal function in
Section~\ref{sec:calcRestA}.

Having obtained the homogeneous solutions of the second-order equation, we
follow Ref.~\cite{Adams:2018yfj} and attempt to find a new set of master
integrals $J_{30}',J_{31}'$ with the help of the following ansatz
\begin{equation}
  \begin{split}
    J_{30}^{\prime}
    &= \frac{J_{30}}{\psi} \ep^2 , 
      \;\;\;
      J_{31}^{\prime} = C_1(x,y) \frac{\psi^2}{\ep} ( \partial_x  J_{30}^{\prime} ) + C_2(x,y) \psi^2 J_{30}^{\prime},
  \end{split}
  \label{eq4.17}
\end{equation}
where $\psi$ is one of the solutions of Eq.~\eqref{eq:elOneDE} that we take to
be $\psi = K(\lambda)/x(1-x y)$. We note that this equation is a linear map
\begin{equation}
  \left ( 
    \begin{array}{c}
      J_{30}' \\
      J_{31}'
    \end{array}
  \right )
  = \hat T 
  \left ( 
    \begin{array}{c}
      J_{30} \\
      J_{31}
    \end{array}
  \right ),
  \label{eq4.18}
\end{equation}
where $\hat T$ is a two-by-two matrix with the zero entry in the upper right
corner. In Eq.~(\ref{eq4.17}), the integral $J_{31}$ is hidden in the term
$\partial_x J_{30}'$. Indeed, if the derivative of $J_{30}'$ is computed
explicitly, the term $\partial_x J_{30}$ appears; we replace this term using
Eq.~(\ref{eq:elDE}) omitting the inhomogeneous term $B \vec J_s$ on the
right-hand side. Since the two coefficients $C_{1,2}$ in Eq.~(\ref{eq4.17}) are
still arbitrary, no information is lost when using this map.

By requiring the $\ep$-form of the new homogeneous block, the unknown
coefficients $C_{1,2}$ can be fixed. We find
\begin{equation}
  \begin{split}
    C_1(x,y) &= x^2 (x - 1) (x + 1) (x - y) (x y - 1), \\
    C_2(x,y) &= x^4 y^2 - 2 x^2 y^2 - 2 x^3 y + 2 x^4 - 3 x^2 + 4 x y.
  \end{split}
\end{equation}
The  matrix $T$ in  Eq.~(\ref{eq4.18}) evaluates to 
\begin{equation}
\hat T =\left(
\begin{array}{cc}
-\frac{x \ep ^2 (x y-1)}{K(\lambda )} ,& 0 \\
 \frac{x \ep ^2 K(\lambda ) \left(x^2 y^2+2 x^2-6 x y+2 y^2+1\right)}{1-x y}+x \ep  E(\lambda )
   (1-x y), & 2 \ep  K(\lambda ) (x-y) (x y-1) \\
\end{array}
\right),
\end{equation}
where $E(\lambda)$ is the complete elliptic integral of the second type. The
differential equation for the new integrals $J_{30,31}^{\prime}$ becomes
\begin{equation}
  \label{eq:elDEprime}
  \partial_x
  \begin{pmatrix}
    J_{30}^{\prime}\\
    J_{31}^{\prime}
  \end{pmatrix} =
  \ep A_1^{\prime}
  \begin{pmatrix}
    J_{30}^{\prime}\\
    J_{31}^{\prime}
  \end{pmatrix}
  + B^{\prime} \vec{J}_s,
\end{equation}
where $\ep$ is factored  out of the new homogeneous block, and the matrix $A_1^{\prime}$ reads
\begin{equation}
A_1^{\prime} = 
\left(
\begin{array}{cc}
 -\frac{x^3 y^2+2 x^3-2 x^2 y-2 x y^2-3 x+4 y}{(x-1) x (x+1) (x-y) (x y-1)}, & \frac{1}{\psi^2 (x-1) x^2 (x+1) (x-y) (x y-1)} \\
 \frac{\psi^2 x^2 \left(x^2 y^2-2 x^2+2 x y-2 y^2+1\right)^2}{(x-1) (x+1) (x-y) (x y-1)}, & -\frac{x^3 y^2+2 x^3-2 x^2 y-2 x y^2-3 x+4 y}{(x-1) x (x+1)
   (x-y) (x y-1)} \\
\end{array}
\right).
\end{equation}
We note that $A_1'$ contains the function $\psi$ and rational functions of $x,
y$. The same transformation also applies to the second pair of master integrals,
i.e., $J_{34,35}$, bringing their homogeneous block to $\ep$-form.

Having diagonalized  two elliptic blocks,  we obtain the full set of differential equations
\begin{equation}
  \label{eq:DEmatDEfprime}
  \begin{split}
  \partial_x \vec{J}^{\prime}(x,y) &= \left(M_{x,0}^{\prime}+\ep M_{x,1}^{\prime}\right) \vec{J}^{\prime}(x,y), \\
  \partial_y \vec{J}^{\prime}(x,y) &= \left(M_{y,0}^{\prime}+\ep M_{y,1}^{\prime}\right) \vec{J}^{\prime}(x,y), 
  \end{split}
\end{equation}
where matrices $M_{x,0}^{\prime}$ and $M_{y,0}^{\prime}$ are lower-triangular.
It is then possible to bring these systems of equations to an $\ep$-form by
rotating the integral basis with an $\ep$-independent matrix $T$ that satisfies
\begin{equation}
    \partial_x T = M_{x,0}^{\prime} T, \quad \partial_y T = M_{y,0}^{\prime} T.
\end{equation}
We use the  Magnus series method~\cite{Argeri:2014qva} to find this matrix.

After combining all  transformations, the new system of differential equations is in $\ep$-form
\begin{equation}
  \label{eq:epsDE}
  \partial_x \vec{\tilde{J}}(x,y) = \ep \widetilde{M}_x \vec{\tilde{J}}(x,y), \quad
  \partial_y \vec{\tilde{J}}(x,y) = \ep \widetilde{M}_y \vec{\tilde{J}}(x,y).
\end{equation}
The matrices $\widetilde{M}_x$ and $\widetilde{M}_y$ contain
\begin{itemize}
    \item rational functions of $x$ and $y$;
    \item two square roots
    \begin{equation}\label{eq:sqrtR}
        R_1(y) = \sqrt{y^2-1}, \quad
        R_2(x,y) = \sqrt{1 + 8 x^2 - 6 x y + x^2 y^2},
    \end{equation}
    that do not couple to  each other;
    \item the elliptic integral $K(\lambda)$,  and a new function that is defined as an integral over $K(\lambda)$,
    \begin{equation}
    \mathcal{H}(x,y) = \int_{0}^x \dm t \frac{\left(2 t^2 y^2+4 t^2-t y^3-7 t y+3 y^2-1\right) K\left(\frac{\left(1-t^2\right) \left(1-y^2\right)}{(1-t y)^2}\right)}{(t y-1) R_2(t,y)^3}.
\end{equation}
\end{itemize}

We solve the differential equations in Eq.~\eqref{eq:epsDE} in terms of iterated
integrals~\cite{Chen:1977oja}. We find that some of them involve complicated
integration kernels containing square roots and elliptic integrals. However, it
turns out that much of this complexity is superfluous since the results for
individual integrals are much more complex than the final result for the
integrated double-emission eikonal function.

Indeed, we find that all elliptic integrals disappear from the final result for
the integrated massive-massless eikonal function. However, this does not happen
naturally and, to reach this conclusion, it is extremely important to simplify
intermediate and final expressions by carefully removing all linearly-dependent
functions from the final result. Key to such simplifications is an observation
that iterated integrals satisfy shuffle algebra relations, whose implementation
in the \texttt{HarmonicSums} package \cite{Ablinger:2012ufz} was very useful for
the current calculation.

Another ingredient required for the final result are the boundary conditions,
which cannot be determined from the differential equations. To fix them, it is
best to compute required integrals at $\beta=0$, which is the regular point for
all integrals. A further benefit of this point is that the dependence on the
angle $\theta$ disappears at $\beta = 0$, so that one computes constants and not
functions of this angle. We present the needed boundary integrals in
Appendix~\ref{sec:bcCalc}.

The final results for the integrated double-emission eikonal functions are
expressed through iterated integrals defined as follows
\begin{equation}
  \label{eq:IterIntDef}
  I[w_1(t_1),\dots,w_n(t_n)|x] = \int\limits_0^x\dm t_1 w_1(t_1)
  \int\limits_0^{t_1}\dm t_2 w_2(t_2)
  \cdots
  \int\limits_0^{t_{n-1}}\dm t_n w_n(t_n),
\end{equation}
where  
\begin{equation}
  \label{eq:IIabc}
  \begin{split}
  w(t) =& \left\{  \frac{1}{t}, \frac{1}{t\pm 1}, \frac{1}{t-y},  \frac{y}{t y-1}, \frac{y \pm 1}{t(y \pm 1)-2}, \frac{1}{t-(y \pm \sqrt{y^2 - 1})}, \right. \\
    & \qquad\qquad \left.
    \frac{1}{t R_2(t,y)}, \frac{1}{(1\pm t)R_2(t,y)}, \frac{1}{(t- y)R_2(t,y)} \right\}.
   \end{split}
\end{equation}
Iterated integrals without the square root $R_2$ can be expressed through
standard Goncharov polylogarithms (GPLs)~\cite{Goncharov:1998kja,
Goncharov:2001iea} with the argument $x$ and indices drawn from the following
set
\begin{equation}
  \label{eq:Gabc}
  \left\{0, \pm 1, y \pm \sqrt{y^2 - 1}, \frac{2}{y \pm 1}, y, \frac{1}{y}\right\}.
\end{equation}
When the square root $R_2$ is present in an integrand, the situation is more
complex. However, we note that we can rationalize it by expressing $x$ through a
new variable $\eta$ and $y$ in the following way
\begin{equation}
  \label{eq:etaDef}
  x = \frac{2\eta}{\eta^2+6\eta y + 8(y^2-1)}.
\end{equation}
The inverse  transformation  reads
\begin{equation}
    \eta = \frac{1-3x y \pm R_2(x,y)}{x}.
\end{equation}
As the result, the relevant iterated integrals with $R_2$ can also be expressed
through GPLs with the argument $\eta$ and indices drawn from the set
\begin{equation}
\label{eq4.27}
    \left\{0, -2\left(y \pm 1\right),-4\left(y \pm 1\right),\frac{2(1-y^2)}{y},-3y \pm \sqrt{y^2+8}, 2\left(-y \pm \sqrt{2-y^2}\right)\right\}.
\end{equation}

\subsection{Simplification of the final result}
\label{sect:4.3}

The result for the integrated double-emission eikonal function written in terms
of GPLs with arguments $\{\eta, x\}$ and indices drawn from Eqs.~(\ref{eq:Gabc},
\ref{eq4.27}) is not suitable for the high-precision, fast numerical evaluation.
Because of this, we decided to construct an optimized form of the result, by
expressing all GPLs up to weight four through normal logarithms, classical
polylogarithms $\textrm{Li}_n$ ($n=2,3,4$) and an additional function
$\textrm{Li}_{2,2}$, defined as
\begin{equation}
\label{eq:Li22def}
\mathrm{Li}_{2,2}(z_1,z_2) = \sum\limits_{i>j>0}^\infty \frac{z_1^i z_2^j}{i^2 j^2} = 
\sum\limits_{i=1}^\infty \sum\limits_{j=1}^\infty 
 \frac{z_1^i}{(i+j)^2} 
 \frac{(z_1 z_2 )^j}{j^2}.
\end{equation}
To do this, we employ the symbol technique~\cite{Goncharov:2010jf, Duhr:2011zq}
since it provides a systematic way to derive non-trivial relations among GPLs.

We work with the integrated double-emission eikonal function as a whole, since
significant simplifications can only be expected in the full result. We combine
GPLs according to their weights and compute their symbols. From the list of
symbols, we extract independent symbol letters and construct candidates for
arguments of $\textrm{Li}_n$ and $\textrm{Li}_{2,2}$ functions, using the
algorithms described in Ref.~\cite{Duhr:2011zq}.

Since there are too many functions that reproduce symbols of the identified sets
of GPLs, we need to reduce their number by imposing additional conditions on
their arguments. For example, we may try to require that, for physical values of
$\beta$ and $\cos \theta$, arguments of the polylogarithmic functions
$\textrm{Li}_n(z_1)$, $\textrm{Li}_{2,2}(z_2, z_3)$ satisfy
\begin{equation}
\label{eq:convergeRegionOfLi}
  {\rm Im} z_{1,2,3} = 0,\quad  | z_1 | \leq 1, \quad | z_2 | \leq 1, \quad | z_2 z_3 | \leq 1.
\end{equation}
If this can be achieved, all functions become real-valued, and can be computed
using convergent series expansion, leading to fast and efficient numerical
evaluation. We note that when testing conditions in
Eq.~\eqref{eq:convergeRegionOfLi}, we choose a particular solution to
Eq.~\eqref{eq:etaDef}
\begin{equation}
    \eta = \frac{1-3x y -R(x,y)}{x}.
\end{equation}

Unfortunately, it turns out that conditions in Eq.~(\ref{eq:convergeRegionOfLi})
are too restrictive. In fact, we find that arguments of all $\textrm{Li}_{2,2}$
functions can be chosen to satisfy Eq.~(\ref{eq:convergeRegionOfLi}). However,
for polylogarithms $\textrm{Li}_n$, it is only possible to choose argument $z_1$
that are smaller than one, i.e. $z_1 < 1$, as opposed to $|z_1|< 1$.
Furthermore, we have to allow for the possibility that $z_1$ is \emph{complex}.
For such cases, we choose
\begin{equation}
    \textrm{Li}_n(z_1) + \textrm{Li}_n(z^*_1), \quad
    \left[\textrm{Li}_n(z_1) - \textrm{Li}_n(z^*_1)\right]/i,
\end{equation}
as suitable candidate functions.

Proceeding along these lines, we rewrite the original result for the integrated
double-emission eikonal functions in terms of polylogarithmic functions with the
chosen properties at the symbol level. The transformed and original expressions
have identical symbols but are not yet the same, because of the missing
polylogarithmic functions of lower weights multiplied by powers of $\pi$ or
$\zeta_3$, and constant terms. Such terms can be restored by applying the
so-called co-product formalism \cite{Goncharov:2005sla, Duhr:2012fh}. By
computing various co-products of the difference between the original and
transformed results, one obtains symbols of a lower weight that can be
reconstructed using the same functional basis. Finally, to fix the constants, we
numerically evaluate the difference between the original and the reconstructed
result, and use PSLQ algorithm ~\cite{pslq} to express it through irrational
constants of various weights, summarized in Table~\ref{tab: constantGPLs}. The
motivation to consider these quantities comes from the expansion of the final
result at small $\beta$, see Eq.~\eqref{eq: betaExpansion}.
\begin{table}[t]
    \centering
    \begin{tabulary}{1.0\textwidth}{|C|C|C|C|C|}
        \hline
        weight & 1 & 2 & 3 & 4 \\ \hline
        number(s)   & $0$   & $\pi^2$   & $\zeta_3, \pi^2 \log(2)$  & $\textrm{Li}_4\left(\frac{1}{2}\right)+\frac{\log^4(2)}{24}, \pi^4, \zeta_3 \log(2), \pi^2 \log^2(2)$   \\ \hline
    \end{tabulary}
    \caption{Pure constants that are needed in the final step of the  simplification procedure.}
    \label{tab: constantGPLs}
\end{table}
We have checked that the old and new expressions agree with each other at
multiple test points in the region $x \in [0,1], y\in [-1,1]$. We note that
throughout the process of manipulating GPLs, computing symbols and co-products,
we heavily relied on the package \texttt{PolyLogTools}~\cite{Duhr:2019tlz,
Maitre:2005uu, Maitre:2007kp} and used the library
\texttt{GiNaC}~\cite{Bauer:2000cp, Vollinga:2004sn} to evaluate GPLs
numerically.

 The integrated double-emission eikonal functions are expressed in terms of
conventional polylogarithms and ${\rm Li}_{2,2}$. This is extremely useful
since, with the new representation, it takes about a \emph{second} to obtain a
numerical value for $\intSS[\tilde S_{ij}]$ and $\intSS[\tilde I_{ij}]$ for a
generic phase space point in \texttt{Mathematica}, as opposed to \emph{minutes}
when using the original expressions in terms of GPLs. The final results for
$\intSS[\tilde S_{ij}]$ and $\intSS[\tilde I_{ij}]$ can be found in ancillary
files provided with this paper. There, a C code for fast and reliable evaluation
of these quantities can be found, which uses algorithms described in
Ref.~\cite{Frellesvig:2016ske} to evaluate all the polylogarithmic functions. It significantly improves the efficiency of
numerical evaluation, especially at the phase-space points where $\beta \sim 1$,
and delivers results for a generic phase-space point in \emph{milliseconds}.

We conclude this discussion by presenting the final results for the integrated
double-emission eikonal functions up to order $\mathcal{O}\left(\ep^0\right)$.
We find
\begin{align}
    \intSS\left[ \widetilde{\mathcal{S}}_{ij} \right] &= \frac{-\Nep^2}{4\ep\emax^{4\ep}} 
    \Bigg\{
    -\frac{1}{4\ep^3}
    +\frac{1}{\ep^2}
    \left[
        \frac{1}{24}+\frac{1}{2}\log\left(\frac{(1-x y)^2}{1-x^2}\right)
    \right]
  \nonumber   \\
    & + \frac{1}{\ep}
    \Bigg[
        \frac{119}{36}
        -\frac{13}{12} \log\left(\frac{(1-x y)^2}{1-x^2}\right)+\frac{\log
   \left(\frac{1-x}{x+1}\right)}{x}-\frac{11 \log (2)}{6} \nonumber \\
   &
   -2 \text{Li}_2\left(\frac{x (y-1)}{1-x}\right)-2 \text{Li}_2\left(\frac{x (y+1)}{x+1}\right) 
   -\frac{1}{2} \log^2\left(\frac{(1-x y)^2}{1-x^2}\right)
   -\frac{\pi
   ^2}{24}
    \Bigg] 
    \label{eq: resultSij} \\
    & + 
    \Bigg[
    -\frac{589}{108}
    -\frac{31}{9} \log \left(\frac{(1-x y)^2}{1-x^2}\right)-\frac{19 \log
   \left(\frac{1-x}{x+1}\right)}{6 x}+\frac{269 \log (2)}{36} \nonumber \\
   &
   +\frac{f_{1}}{x}-\frac{f_{2}}{12}+\frac{11}{3} \log (2) \log \left(\frac{(1-x y)^2}{1-x^2}\right)+\frac{\pi
   ^2}{12}+\frac{11 \log ^2(2)}{6} 
   \nonumber \\
   &
   -\frac{f_{3}}{3}+\frac{1}{12} \pi ^2 \log \left(\frac{(1-x y)^2}{1-x^2}\right)-\frac{11
   \zeta_3}{8}
    \Bigg]
    +\mathcal{O}\left(\ep\right)
    \Bigg\},
    \nonumber 
\end{align}
and 
\begin{equation}
\label{eq: resultIij}
\begin{split}
    \intSS\left[ \widetilde{\mathcal{I}}_{ij} \right] &= \frac{-\Nep^2}{4\ep\emax^{4\ep}} 
    \Bigg\{
    -\frac{1}{12\ep^2}
    +\frac{1}{\ep}
    \left[
    \frac{25}{72}
    +\frac{1}{6} \log \left(\frac{(1-x y)^2}{1-x^2}\right)-\frac{\log (2)}{3}
    \right] \\
    &
    +\Bigg[
    -\frac{131}{216}
    -\frac{37}{36} \log \left(\frac{(1-x y)^2}{1-x^2}\right)+\frac{\log \left(\frac{1-x}{x+1}\right)}{3 x}+\frac{59 \log
   (2)}{36} \\
   &
    -\frac{2}{3} \text{Li}_2\left(\frac{x (y-1)}{1-x}\right)-\frac{2}{3} \text{Li}_2\left(\frac{x (y+1)}{x+1}\right)-\frac{1}{6} \log
   ^2\left(\frac{(1-x y)^2}{1-x^2}\right) \\
   & 
   +\frac{2}{3} \log (2) \log \left(\frac{(1-x y)^2}{1-x^2}\right)+\frac{\log^2(2)}{3}
    \Bigg]
    +\mathcal{O}\left(\ep\right)
    \Bigg\},
\end{split}
\end{equation}
where we used the abbreviations
\begin{align}
    f_{1} =& -\text{Li}_2\left(\frac{x (y-1)}{1-x}\right)+\text{Li}_2\left(\frac{x (y+1)}{x+1}\right)-4 \text{Li}_2\left(\frac{1-x}{2}\right)+4 \text{Li}_2(-x)-4
   \text{Li}_2(x) \\ 
   &+\log \left(\frac{1-x}{x+1}\right) \log \left(\frac{1-x y}{x+1}\right)-2 \log ^2\left(\frac{1-x}{2}\right)+\frac{1}{2} \log
   ^2\left(\frac{1-x}{x+1}\right)+\frac{\pi ^2}{3}, \nonumber\\
   f_{2} =& -40 \text{Li}_2\left(\frac{x (y-1)}{1-x}\right)-40 \text{Li}_2\left(\frac{x (y+1)}{x+1}\right)-20 \log ^2\left(\frac{1-x y}{1-x}\right) \\
   &-20 \log^2\left(\frac{1-x y}{x+1}\right) +13 \log ^2\left(\frac{1-x}{x+1}\right) \nonumber\\
   f_{3} =& 18 \text{Li}_3\left(\frac{x (y+1)}{x+1}\right)+12
   \text{Li}_3\left(\frac{1-x}{1-x y}\right)-18 \text{Li}_3\left(\frac{x
   (1-y)}{1-x y}\right) \\
   &+12 \text{Li}_3\left(\frac{(x+1) (1-y)}{2 (1-x
   y)}\right)+12 \text{Li}_3\left(\frac{(1-x) (y+1)}{2 (1-x y)}\right)+30
   \text{Li}_3\left(\frac{1-x y}{x+1}\right)\nonumber\\
   &-12\text{Li}_3\left(\frac{1-x}{x+1}\right)-12
   \text{Li}_3\left(\frac{1-y}{2}\right)-12
   \text{Li}_3\left(\frac{y+1}{2}\right) \nonumber\\
   &+\log (1-x) \Big[-27 \text{Li}_2\left(\frac{1-x}{2}\right)+33
   \text{Li}_2(1-x)+27 \text{Li}_2(-x)+6 \text{Li}_2(x) 
   \nonumber\\ 
   &+\frac{39}{2}
   \text{Li}_2\left(\frac{1-y}{2}\right)+\frac{15}{2}
   \text{Li}_2\left(\frac{2 x (y+1)}{(x+1)^2}\right)+12
   \text{Li}_2\left(\frac{x (y+1)}{x+1}\right)
   \nonumber\\
   &-15
   \text{Li}_2\left(-\frac{1-x^2}{x^2-2 y x+1}\right)-\frac{15}{2}
   \text{Li}_2\left(\frac{2 x (1-y)}{x^2-2 y x+1}\right)+15
   \text{Li}_2\left(\frac{x (x+1) (1-y)}{x^2-2 y x+1}\right)
   \nonumber\\
   &+\frac{15}{2}
   \text{Li}_2\left(\frac{(1-x)^2 (y+1)}{2 \left(x^2-2 y
   x+1\right)}\right)-33 \text{Li}_2\left(\frac{1-x}{1-x
   y}\right)\Big]
   \nonumber\\
   &+\log (1-x y) \Big[15
   \text{Li}_2\left(\frac{1-x}{2}\right)-42 \text{Li}_2(1-x)+12
   \text{Li}_2(-x)-27 \text{Li}_2(x)
   \nonumber\\
   &-27
   \text{Li}_2\left(\frac{1}{x+1}\right)-\frac{15}{2}
   \text{Li}_2\left(\frac{1-y}{2}\right)-\frac{15}{2}
   \text{Li}_2\left(\frac{2 x (y+1)}{(x+1)^2}\right)+6
   \text{Li}_2\left(\frac{x (y+1)}{x+1}\right)
   \nonumber\\
   &+15
   \text{Li}_2\left(-\frac{1-x^2}{x^2-2 y x+1}\right)+\frac{15}{2}
   \text{Li}_2\left(\frac{2 x (1-y)}{x^2-2 y x+1}\right)-15
   \text{Li}_2\left(\frac{x (x+1) (1-y)}{x^2-2 y x+1}\right)
   \nonumber\\
   &-\frac{15}{2}
   \text{Li}_2\left(\frac{(1-x)^2 (y+1)}{2 \left(x^2-2 y
   x+1\right)}\right)+21 \text{Li}_2\left(\frac{1-x}{1-x
   y}\right)\Big]
   \nonumber\\
   &+\log (x+1) \Big[27
   \text{Li}_2\left(\frac{1-x}{2}\right)-9 \text{Li}_2(1-x)-21
   \text{Li}_2(-x)+3 \text{Li}_2(x)+9
   \text{Li}_2\left(\frac{1}{x+1}\right)
   \nonumber\\
   &-12
   \text{Li}_2\left(\frac{1-y}{2}\right)-18 \text{Li}_2\left(\frac{x
   (y+1)}{x+1}\right)-15 \text{Li}_2\left(\frac{2 x (1-y)}{x^2-2 y
   x+1}\right)+15 \text{Li}_2\left(\frac{x (x+1) (1-y)}{x^2-2 y
   x+1}\right)
   \nonumber\\
   &-3 \text{Li}_2\left(\frac{1-x}{1-x y}\right)+15
   \text{Li}_2\left(\frac{1-x^2}{2 (1-x y)}\right)\Big]
   \nonumber\\
   &+\log (2)
   \Big[-\frac{45}{2} \text{Li}_2\left(\frac{1-x}{2}\right)+\frac{45}{2}
   \text{Li}_2\left(\frac{2 x (1-y)}{x^2-2 y x+1}\right)-\frac{45}{2}
   \text{Li}_2\left(\frac{x (x+1) (1-y)}{x^2-2 y x+1}\right)
   \nonumber\\
   &-\frac{45}{2}
   \text{Li}_2\left(\frac{1-x^2}{2 (1-x y)}\right)-\frac{45}{2}
   \text{Li}_2\left(\frac{x (1-y)}{1-x y}\right)\Big]
   \nonumber\\
   &+\log (1-y)
   \Big[\frac{15}{2} \text{Li}_2\left(\frac{1-x}{2}\right)-\frac{15}{2}
   \text{Li}_2\left(\frac{2 x (1-y)}{x^2-2 y x+1}\right)+\frac{15}{2}
   \text{Li}_2\left(\frac{x (x+1) (1-y)}{x^2-2 y x+1}\right)
   \nonumber\\
   &+\frac{15}{2}
   \text{Li}_2\left(\frac{1-x^2}{2 (1-x y)}\right)+\frac{15}{2}
   \text{Li}_2\left(\frac{x (1-y)}{1-x y}\right)\Big]
   \nonumber\\
   & -6 \log ^2(1-x) \log (1-y)+6 \log ^2(1-x) \log (1-x y)-\frac{15}{2} \log
   (1-x) \log ^2(1-x y)
   \nonumber\\
   &-\frac{1}{2} \log ^3(1-x y)+9 \log (x+1) \log
   ^2(1-x y)-\frac{3}{2} \log (1-y) \log ^2(1-x y)
   \nonumber\\
   &+21 \log (y+1) \log
   ^2(1-x y)+21 \log ^2(x+1) \log (y+1)-\frac{9}{2} \log ^2(x+1) \log
   (1-x y)
   \nonumber\\
   &+\frac{9}{2} \log (x+1) \log (1-x) \log (1-y)+\frac{39}{2} \log
   (1-x) \log (1-y) \log (y+1)
   \nonumber\\
   &-39 \log (x+1) \log (1-x) \log (1-x
   y)+\frac{15}{2} \log (1-x) \log (1-y) \log (1-x y)
   \nonumber\\
   &+3 \log (x+1) \log
   (1-y) \log (1-x y)-42 \log (x+1) \log (y+1) \log (1-x y)
   \nonumber\\
   &-\frac{39}{2}
   \log (1-y) \log (y+1) \log (1-x y)+\log ^3(1-x)+12 \log (x+1) \log
   ^2(1-x)
   \nonumber\\
   &+39 \log ^2(x+1) \log (1-x)-\frac{19}{2} \log ^3(x+1) \nonumber\\
   &+ \log ^2(2) \left[-\frac{69}{2} \log (1-x y)+51 \log (1-x)+\frac{87}{2}
   \log (x+1)+\frac{15}{2} \log (1-y)\right]
   \nonumber\\
   &+\log (2) \Big[-\frac{69}{2}
   \log ^2(1-x y)-\frac{69}{2} \log (1-x) \log (1-y)-\frac{39}{2} \log
   (1-x) \log (y+1)
   \nonumber\\
   &+\frac{99}{2} \log (1-x) \log (1-x y)-\frac{15}{2}
   \log (x+1) \log (1-y)+42 \log (x+1) \log (1-x y)
   \nonumber\\
   &+27 \log (1-y) \log
   (1-x y)+\frac{39}{2} \log (y+1) \log (1-x y)-15 \log ^2(1-x)
   \nonumber\\
   &-27 \log
   ^2(x+1)-\frac{135}{2} \log (x+1) \log (1-x)\Big]
   \nonumber\\
   &+\pi ^2
   \left[\frac{33}{4} \log (1-x y)-\frac{9}{4} \log (1-x)-\frac{5}{4}
   \log (1-y)\right] \nonumber\\
   &-30 \zeta_3-\frac{45 \log^3(2)}{2}+\frac{15}{4} \pi ^2 \log (2).
   \nonumber 
\end{align}
For reference, we present numerical results for a few benchmark points in
Table~\ref{tab: benchmark}.
\begin{table}[t]
\resizebox{1.0\textwidth}{!}{
\begin{tabular}{|c|cccc|ccc|}\hline
\multirow{2}{*}{$(x, y)$} & \multicolumn{4}{c|}{$\intSS\left[ \widetilde{\mathcal{S}}_{ij} \right]$} & \multicolumn{3}{c|}{$\intSS\left[ \widetilde{\mathcal{I}}_{ij} \right]$} \\ \cline{2-8}
                       & $\ep^{-2}$   & $\ep^{-1}$   & $\ep^{0}$   & $\ep^{1}$   & $\ep^{-1}$      & $\ep^{0}$     & $\ep^{1}$     \\ \hline
 $(0.1, 0.2)$  & $0.0264891$  &  $-0.40119472$  &  $0.066258771$  & $2.1059932$  &  $0.11111398$  & $0.020954787$  &  $0.083825324$     \\
 $(0.9, 0.1)$  & $0.7777216$  &  $1.8762882$	   &  $6.6664261$    & $17.203106$  &  $0.36152480$	 & $0.47275868$	  &  $1.0676022$        \\
 $(0.8, 0.9)$  & $-0.7204734$ &  $-2.2382469$   &  $-4.0335936$	 & $-4.3260204$ &  $-0.13787352$ & $-0.28578587$  &  $-0.35505627$      \\
 $(0.8, -0.9)$ & $1.0948166$  & $1.4353310$	   & $3.4641280$	 & $8.3534044$  &  $0.46722313$	 & $0.19603765$	  &  $0.36091849$ \\ \hline
\end{tabular}
}
\caption{Benchmark points for the integrated   double-emission eikonal functions. The prefactor $-1/(4\ep\emax^{4\ep})\Nep^2$ is not included. The highest poles do not depend on $x$ and $y$ are not shown, c.f.  Eqs.~(\ref{eq: resultSij}, \ref{eq: resultIij}).}
\label{tab: benchmark}
\end{table}

\subsection{Checks of the result }

We have performed several checks of the obtained results at various stages of
the calculation. For example, we have computed master integrals numerically
using dimensional shifts, see Eq.~(\ref{eq:drr}), and direct integration as
described in Appendix~\ref{sec:directInt}. To check complete results for
$\intSS[\tilde S_{ij}]$ and $\intSS[\tilde I_{ij}]$, we recalculated them using
the parameterization of the Heaviside function in Eq.~(\ref{eq:thetaInt}). Since
the Heaviside function is mapped onto a delta-function, a conventional reduction
to master integrals with the help of publicly-available codes becomes possible.
However, since we only need to check the calculation described above, we decided
to compute its small-$\beta$ expansion, which is much easier to do. We then
compare it to the results described in Section~\ref{sect:4.3} , which we also
expand around $\beta=0$. Perfect agreement up to order $\beta^4$ is found for
the two results. It is also straightforward to extend the comparison to higher
orders in the $\beta$-expansion. For completeness, we present the leading term
in the $\beta$-expansion for the two integrated double-emission eikonal
functions
\begin{equation}
\label{eq: betaExpansion}
\begin{split}
   &  \intSS\left[ \widetilde{\mathcal{S}}_{ij} \right] \Bigg |_{\beta = 0} = \frac{-\Nep^2}{4\ep\emax^{4\ep}} \Bigg[
    -\frac{1}{4 \ep ^3}
    +\frac{1}{24 \ep ^2}
    +\frac{1}{\ep } \left(\frac{47}{36}-\frac{11 \log (2)}{6}-\frac{\pi ^2}{24}\right)
    \\
   & + \left(-\frac{553}{108}+\frac{269 \log(2)}{36}+\frac{\pi ^2}{12}+\frac{11 \log ^2(2)}{6}-\frac{11 \zeta _3}{8}\right)
   + \ep  \Bigg (
   \frac{3887}{324}
   \\
   & 
   -\frac{1133 \log (2)}{54}
   -\frac{47 \pi ^2}{144}
   -\frac{341 \log ^2(2)}{36}
   +\frac{99 \zeta _3}{8}
   +\frac{11}{36} \pi ^2 \log (2)
     -\frac{11 \log^3(2)}{9}  \\
   & 
    -\frac{29 \pi ^4}{1440}
   -2 \text{Li}_4\left(\frac{1}{2}\right)
   -\frac{\log ^4(2)}{12}
   -\frac{7}{4} \zeta _3 \log (2)
   +\frac{1}{12} \pi ^2 \log ^2(2)
   \Bigg )
  +O\left(\ep ^2\right )
\Bigg], \\
 &   \intSS\left[ \widetilde{\mathcal{I}}_{ij} \right] \Bigg |_{\beta = 0} = \frac{-\Nep^2}{4\ep\emax^{4\ep}}  
   \Bigg[
   -\frac{1}{12 \ep ^2}
   +\frac{1}{\ep }\left(\frac{25}{72}-\frac{\log (2)}{3}\right)
 \\
   & +\left(-\frac{275}{216}+\frac{59 \log (2)}{36}+\frac{\log ^2(2)}{3}\right) 
    +\ep  \left(
   \frac{2737}{648}
   -\frac{577 \log (2)}{108}
   -\frac{23 \pi^2}{144}
   -\frac{95 \log ^2(2)}{36}
   \right. \\
   & \left. \qquad \qquad \qquad \qquad
   +\frac{7 \zeta _3}{4}
   +\frac{1}{18} \pi ^2 \log (2)
   -\frac{2 \log ^3(2)}{9}
   \right)+O\left(\ep ^2\right)
   \Bigg].
\end{split}
\end{equation}
Finally, in Section~\ref{sec:calcRestA} we describe an alternative computation
of the integrated double-emission eikonal functions that does not involve IBP
reduction, master integrals and differential equations. The agreement between
calculations described in this section and in section~\ref{sec:calcRestA}
provides a very strong check on their correctness.

\section{An alternative computation in the massive parton's rest frame}
\label{sec:calcRestA}

The final result for the integrated double-emission eikonal function in the
two-gluon case, derived in the previous sections, is simpler than results for
individual integrals. This is certainly true for the divergent contributions,
but also for the finite part since elliptic integrals disappear from the final
expression. It is interesting to understand the origin of this simplicity. A
possible reason to expect it is Lorentz invariance, i.e. the possibility to
choose a suitable reference frame (for example, the rest frame of the massive
parton) to compute the integrated double-emission eikonal function.

Lorentz invariance would have been very useful for computing integrated
double-emission eikonal functions, if not for the Heaviside functions that
appear because of how subtraction terms in the nested soft-collinear subtraction
scheme are defined. These Heaviside functions depend on the energies of
unresolved partons in the laboratory frame and, therefore, transform in a
non-trivial way under Lorentz boosts. Nevertheless, as we explain below it turns
out to be very useful to compute the integrated eikonal in the rest frame of the
massive parton, as it allows us to understand the simplicity of the result, at
least partially.

The integrated double-emission eikonal function defined in
Eq.~\eqref{eq:ssintNoThKtoL} is written in a Lorentz-invariant way. In the
computation reported in the previous sections, we have taken $P = (1, \vec 0)$
defining the laboratory frame, but this is certainly not the only option. In
fact, we find it useful to consider a frame where the massive parton is at rest
$p_i = (m, \vec 0)$. In this frame, $P=\gamma_i (1,\beta \vec{n}_P)$, where
$\beta$ is the velocity of the massive parton in the lab frame, $\gamma_i =
E_i/m$, where $E_i$ is the energy of parton $i$ in the lab frame, and
$\vec{n}_P$ is $-\vec n_i$, where $\vec n_i$ is defined in the lab frame. The
scalar products evaluate to
\begin{equation}
  l_\xa \cdot P
  = l_\xa^0 \;\gamma_i \;  \rho_{\xa t}, 
  \quad l_\yb \cdot P = l_\yb^0  \; \gamma_i \; 
  \rho_{\yb t},
\end{equation}
where $\rho_{\xa (\yb) t} = 1 - \beta \vec n_P \cdot \vec n_{\xa(\yb)}$.
Furthermore, writing $l_\yb^0 = \omega l_\xa^0$, using the homogeneity of
$\Xi_{ij}$, and integrating over $l_\xa^0$, we obtain the following
representation for the double-soft integral
\begin{equation}
  \label{eq:boostFrameSS}
  \intSS\left[ \Xi_{ij} \right] = 
  \normA \left \langle
  \int \limits_{0}^{\infty}
  \frac{{\rm d} \omega}{\omega^{1+2\ep} } \; \psi_{t, \xa j}^{4 \ep} \;
  \theta( \psi_{t, \xa j }  - \omega \;\psi_{t, \yb j }  )
  \left [ \omega^2 \; \Xi_{ij}(\xa,\yb) \right ] \right \rangle_{\xa \yb}.
\end{equation}
The brackets $\langle .. \rangle_{\xa \yb}$ denote integration over directions
of partons $\xa$ and $\yb$. We note that in the above equation we have defined
the function $\psi_{t,yz} = \rho_{yt}/\rho_{zt}$, and the normalization factor
\begin{equation}
  \label{eq:pAnorm}
  \normA = -\frac{1}{4 \ep} \left ( \frac{\emax}{ \gamma_i \rho_{t j} }  \right )^{-4 \ep}
  = -\frac{1}{4 \ep \emax^{4\ep}} \frac{(1-\beta^2)^{2\ep}}{(1-\beta \cos \theta )^{4\ep}}.
\end{equation}
We also note that the double-emission eikonal function $\Xi_{ij}$ in
Eq.~(\ref{eq:boostFrameSS}) has to be evaluated in the rest frame of parton $i$.
Furthermore, the four-momenta of partons $\xa, \yb$ in that equation should be
taken to be $l_\xa= (1, \vec n_\xa)$, $l_\yb = \omega (1 , \vec n_\yb)$. In the
rest frame of $i$, scalar products $p_i \cdot l_\xa$, $p_i \cdot l_\yb$ become
simple. Apart from the prefactor ${\cal N}_A$, the non-trivial dependence on the
angle between $\vec P$ and $\vec n_j$ resides in functions $\psi_{t,\xa j}$ and
$\psi_{t,\yb j}$, which appear in Eq.~(\ref{eq:boostFrameSS}) either in an
$\ep$-dependent power or inside the Heaviside function. Our goal is to exploit
this fact and write the integral in Eq.~\eqref{eq:boostFrameSS} as a sum of two
terms, one that contains no divergences and can be calculated in \emph{four}
dimensions, and another one which is composed of simpler integrals, can be
calculated analytically in a systematic way and contains all $1/\ep$ poles. Such
a separation, and in particular an analytic form of a divergent term, is very
useful for establishing the explicit cancellation of infra-red poles in the
context of NNLO calculations, see e.g. Ref.~\cite{Devoto:2023rpv}.

 We will focus on the case where soft partons are gluons, since it is more
general than the case where a $q \bar q$ pair is soft. Denoting the integrated
eikonal function for the two gluons as $G_{ij} = \intSS\left[ \tilde{S}_{ij}
\right]$, we proceed with iterative subtraction of singularities, starting with
the soft one. We write
\begin{equation}
  G_{ij} = S_\omega [ G_{ij}] + \bar S_{\omega} [G_{ij}],
\end{equation}
where $\bar S_{\omega} = 1 - S_{\omega}$ and the operator $S_\omega$, acting on
relevant quantities, retains the leading singular contribution in the $\omega
\to 0$ limit. Computation of the strongly-ordered integral $S_\omega[G_{ij}]$ is
discussed in Section~\ref{sec:soLim}. The soft-subtracted integral\footnote{Note
that we do not act with the operator $S_\omega$ on the theta-function.}
\begin{equation}
  \label{eq:soSubtrInt}
  \bar S_\omega \left [ G_{ij} \right ]
  = \normA
  \left \langle
    \int \limits_{0}^{\infty}
    \frac{{\rm d} \omega}{\omega^{1+2\ep} } \psi_{t, \xa j }^{4 \ep}
    \theta( \psi_{t, \xa j}  - \omega \psi_{t, \yb j}   )
    \bar S_\omega \left [ \omega^2 \tilde S_{ij}(\xa,\yb) \right ] \right \rangle_{\xa \yb},
\end{equation}
requires further manipulations.

For reasons that will become clear shortly,  it is convenient to define a new quantity
\begin{equation}
  \bar S_\omega [ \Delta G_{ij} ]
  = \bar S_\omega [G_{ij}] - \bar S_\omega[G^{(0)}_{ij}],
\end{equation}
where 
\begin{equation}
  \bar S_\omega \left [ G^{(0)}_{ij} \right ]
  = \normA
  \left \langle
    \int \limits_{0}^{\infty}
    \frac{{\rm d} \omega}{\omega^{1+2\ep} } 
    \theta( 1 - \omega   )
    \bar S_\omega \left [ \omega^2 \tilde S_{ij}(\xa,\yb) \right ] \right \rangle_{\xa \yb}.
  \label{eq5.8}
\end{equation}
Apart from the $\beta$-dependence in the normalization factor $\normA$, the
above quantity corresponds to the double-soft integral in the laboratory frame
where the massive parton $i$ is now at rest, i.e. $\beta = 0$. If the emitter
$i$ is at rest, the dependence on the emission angle disappears; therefore, the
integral in Eq.~(\ref{eq5.8}) is a number (as opposed to a function), and can be
easily computed. Hence, we write
\begin{equation}
  \bar S_\omega [G_{ij}]
  = \bar S_\omega [\Delta G_{ij}]
  + \bar S_\omega [  G_{ij}^{(0)}],
\end{equation}
and focus on the first term on the right-hand side in what follows. 

Our goal is to remove all divergences from integrands in Eqs.
(\ref{eq:soSubtrInt}, \ref{eq5.8}). These divergences are collinear, and one can
check that the soft-subtracted integrand is singular in only two cases, the
triple-collinear one $\xa || \yb || j$, and the double-collinear one $\xa ||
\yb$. Note that the soft-subtracted double-emission eikonal is \emph{not}
singular in two other possible double-collinear limits $\xa || j$ and $\yb ||
j$, which simplifies the construction of the subtraction terms. To extract
singularities, we write
\begin{equation}
  \begin{split}
    \label{eq:barS-subtr}
    \bar S_\omega  \left [ \Delta G_{ij} \right ]
    & =\normA
      \Big \langle
      \int \limits_{0}^{\infty}  \frac{{\rm d} \omega}{\omega^{1+2\ep} }
      \Big [
      C_{\xa \yb}
      + \bar C_{\xa \yb}  
      \left ( C_{j \xa \yb} + \bar C_{j \xa \yb} \right )
      \Big ]
    \\
    & \times  \left ( 
      \psi_{t, \xa j}^{4 \ep} \theta( \psi_{t, \xa j }  - \omega \psi_{t,  \yb j}  ) - \theta(1-\omega)
      \right )
      \bar S_\omega \left [ \omega^2 \tilde S_{ij}(\xa,\yb) \right ]  \Big \rangle_{\xa \yb},
  \end{split}
\end{equation}
where operators $C_{\xa \yb}$ and $C_{j \xa \yb}$ extract particular singular
limits from all expressions that appear to the right of them, and $\bar C_{\xa
\yb} = 1 - C_{\xa \yb}$, $\bar C_{j \xa \yb} = 1 - C_{j \xa \yb}$. The operator
$C_{\xa \yb}$ extracts the double-collinear limit $\xa || \yb$ and the operator
$C_{j \xa \yb}$ -- the triple- collinear limit $\xa || \yb || j$.\footnote{We
note that we do not restrict action of $C_{\xa \yb}$ to parts of the unresolved
phase space. As we will see later, it is not necessary to do this in the current
case.} An important simplification in Eq.~(\ref{eq:barS-subtr}) occurs because
\begin{equation}
  C_{j \xa \yb} 
  \left ( 
    \psi_{t, \xa j}^{4 \ep} \theta( \psi_{t, \xa j }  - \omega \psi_{t,  \yb j}  ) - \theta(1-\omega)
  \right )
  \bar S_\omega \left [ \omega^2 \tilde S_{ij}(\xa,\yb) \right ] 
  = 0,
  \label{eq5.10}
\end{equation}
which follows from the equations $C_{j \xa \yb} \psi_{t,\xa j} =1 $, $C_{j \xa
\yb} \psi_{t, \yb j} = 1$. The consequence of Eq.~(\ref{eq5.10}) is that the
triple-collinear subtraction in Eq.~(\ref{eq:barS-subtr}) is not needed; hence,
we can replace $C_{j \xa \yb} + \bar C_{j \xa \yb}$ with the identity operator.

 Taking into account this feature, we write 
\begin{equation}
\bar S_\omega \left [ \Delta G_{ij} \right ] = \normA
 \Big \langle
  \int \limits_{0}^{\infty}  \frac{{\rm d} \omega}{\omega^{1+2\ep} }
\left( C + \bar{C} \right)   \bar S_\omega \left [ \omega^2 \tilde S_{ij}(\xa,\yb) \right ] \Big \rangle_{\xa \yb},
\label{eq5.12}
\end{equation}
where operators $C$ and $\bar C$ read
\begin{equation}
  \begin{split}
    C & = 
        \left(\psi_{t, \xa j}^{4 \ep} - 1 \right) \theta(1-\omega) \; C_{\xa \yb} ,\\
    \bar{C} & = \bar C_{\xa \yb}
              \left( \psi_{t, \xa j}^{4 \ep} \theta( \psi_{t, \xa j }  - \omega \psi_{t,  \yb j}  ) - \theta(1-\omega)\right),
  \end{split}
\end{equation}
and we have used $C_{\xa \yb} \psi_{t,\xa j}^{4 \ep} \theta(\psi_{t,\xa j} -
\omega \psi_{t,\yb j}) = \psi_{t, \xa j}^{4 \ep} \theta(1- \omega)$. Since
operator $\bar C$ removes all divergences from the integrand in
Eq.~(\ref{eq5.12}), we can expand it in powers of $\ep$. A convenient first step
is to write $\psi^{4\ep}_{t, \xa j} = 1 + ( \psi^{4\ep}_{t, \xa j} -1 ). $ Upon
doing that, $\bar C$ splits into two terms $\bar C = \bar C_a + \bar C_b$, where
\begin{equation}
  \begin{split}
    \bar{C}_{a} & = \bar C_{\xa \yb} 
                  \left[ \theta( \psi_{t, \xa j }  - \omega \psi_{t,  \yb j}  ) - \theta(1-\omega) \right] = \left[ \theta( \psi_{t, \xa j }  - \omega \psi_{t,  \yb j}  ) - \theta(1-\omega) \right],
    \\
    \bar{C}_{b} & =  \bar C_{\xa \yb}
                  \left(\psi_{t, \xa j}^{4 \ep} - 1 \right)  \theta( \psi_{t, \xa j }  - \omega \psi_{t,  \yb j}  )
    \\
                &  = \left(\psi_{t, \xa j}^{4 \ep} - 1 \right) \left[ \theta( \psi_{t, \xa j }  - \omega \psi_{t,  \yb j}  )  - \theta(1-\omega) C_{\xa \yb}  \right]. 
  \end{split}
\end{equation}
Since the normalization factor ${\cal N}_A$ is ${\cal O}(1/\ep)$ (c.f.
Eq.~(\ref{eq:pAnorm})), and the operator $\bar{C}_{b}$ is $\mathcal{O}(\ep)$, as
follows from the expansion of $\psi^{4 \ep}_{t,\xa j}$ in powers of $\ep$, it
only contributes to ${\cal O}(\ep^0)$ part of the final result. Hence, it can be
calculated numerically right away by integrating in four dimensions.

On the contrary, the integral that involves the operator $\bar{C}_{a}$ requires
higher-order expansion in $\ep$ as it is multiplied with ${\cal N}_A \sim 1/\ep$
and the corresponding integrand does not possess explicit $\ep$-suppression
factors. Hence, we need to simplify it further.

The key observation that allows us to do this is that this integral can be
written in two complementary ways, as follows from the fact that the
soft-subtracted double eikonal function is (nearly) invariant under a combined
$\omega \to 1/\omega$ and $\xa \leftrightarrow \yb$ transformation. The relevant
equation reads
\begin{equation}
  \bar S_\omega \left [ \omega^2 \tilde S_{ij}(\xa,\yb) \right ]
  \Bigg |_{\substack{\omega \to 1/\omega\\ \xa \leftrightarrow \yb}}
  =
  \bar S_\omega \left [ \omega^2 \tilde S_{ij}(\xa,\yb) \right ]
  + \frac{(\rho_{\xa \yb} - \rho_{\xa j} - \rho_{\yb j})
    (\rho_{\xa j} - \rho_{\yb j}
    ) }{\rho_{\xa \yb} \rho_{\xa j} \rho_{\yb j}},
\label{eq5.18}
\end{equation}
and we note that the second term on the right-hand side of the above equation is
much simpler than the complete double-emission eikonal function. To make use of
Eq.~(\ref{eq5.18}), we write
\begin{equation}
  \begin{split}
    I_{a}
    & =
      \normA \Big \langle
      \int \limits_{0}^{\infty}  \frac{{\rm d} \omega}{\omega^{1+2\ep} }
      \bar{C}_{a}
      \bar S_\omega \left [ \omega^2 \tilde S_{ij}(\xa,\yb) \right ] \Big \rangle_{\xa \yb} \\
    & =
      \normA \Big \langle
      \int \limits_{0}^{\infty}  \frac{{\rm d} \omega}{\omega^{1+2\ep} }
      \left[ \theta( \psi_{t, \xa j }  - \omega \psi_{t,  \yb j}  ) - \theta(1-\omega) \right]
      \bar S_\omega \left [ \omega^2 \tilde S_{ij}(\xa,\yb) \right ] \Big \rangle_{\xa \yb}
      \label{eq:Ia}
    \\
    & =  \normA \Big \langle
      \int \limits_{0}^{\infty}  \frac{{\rm d} \omega}{\omega^{1-2\ep} }
      \left[ \theta( \omega \psi_{t,  \yb j} - \psi_{t, \xa j }  ) - \theta(\omega-1) \right]
      \bar S_\omega \left [ \omega^2 \tilde S_{ij}(\xa,\yb) \right ]\Bigg |_{\substack{\omega\to1/\omega\\\xa \leftrightarrow \yb}} \Big \rangle_{\xa \yb} .
  \end{split}
\end{equation}
We then use Eq.~(\ref{eq:Ia}) to combine two different representations for the
integral $I_{a}$ and find
\begin{equation}
  I_{a} =
  F_{a,1} + F_{a,2},
\end{equation}
where the two integrals read
\begin{align}
  F_{a,1}
  & =
    \frac{\normA}{2} \Big \langle
    \int \limits_{0}^{\infty}  \frac{{\rm d} \omega}{\omega}
    \left( \omega^{-2\ep} - \omega^{2\ep} \right)
   f_-(\omega, \psi_{t,\xa j},\psi_{t,\yb j})
    \bar S_\omega \left [ \omega^2 \tilde S_{ij}(\xa,\yb) \right ] \Big \rangle_{\xa \yb},
    \label{eq:F2a} \\
   F_{a,2}
  & =  \frac{\normA}{2} \Big \langle
    \int \limits_{0}^{\infty}  \frac{{\rm d} \omega}{\omega^{1-2\ep} }
    f_+(\omega, \psi_{t,\xa j},\psi_{t,\yb j})
    \frac{(\rho_{\xa \yb} - \rho_{\xa j} - \rho_{\yb j})
    (\rho_{\xa j} - \rho_{\yb j}) }{\rho_{\xa \yb} \rho_{\xa j} \rho_{\yb j}}
    \Big \rangle_{\xa \yb},
    \label{eq:deltaF2a}
\end{align}
and the two functions $f_{\mp}$ are defined as
\begin{equation}
  \begin{split}
    & f_-(\omega,\psi_{t, \xa j}, \psi_{t,\yb j}) = \theta( \psi_{t, \xa j }  - \omega \psi_{t,  \yb j}  ) - \theta(1-\omega) ,
    \\
    & f_+(\omega,\psi_{t, \xa j}, \psi_{t,\yb j}) =\theta( \omega \psi_{t,  \yb j} - \psi_{t, \xa j }  ) - \theta(\omega-1). 
  \end{split} 
\end{equation}

The two integrals in Eqs.~(\ref{eq:F2a}, \ref{eq:deltaF2a}) are finite.
Furthermore, since $\left( \omega^{-2\ep} - \omega^{2\ep} \right) = -4\ep
\log{\omega} + \mathcal{O}(\ep^2)$, the integral $F_{a,1}$ only contributes to
the finite part of Eq.~\eqref{eq:boostFrameSS}, and can be immediately
calculated numerically in four dimensions. The second integral $F_{a,2}$ is
sufficiently simple to be computed analytically; we discuss its computation in
Section~\ref{sec:ga}.

The final expression for the integrated double-emission eikonal function therefore reads 
\begin{equation}
  G_{ij} = S_\omega\left[G_{ij} \right] +
  \bar{S}_\omega\left[G_{ij}^{(0)} \right] + \bar{S}_\omega\left[ \Delta G_{ij}
  \right],
  \label{eq5.20}
\end{equation}
where the first and the second terms on the right-hand side are the
strongly-ordered and the $\beta = 0$ contributions. The third term reads
\begin{equation}
  \label{eq:barSwDeltaG}
  \bar S_\omega[\Delta G_{ij}]
  = \bar S_\omega[\Delta G_{ij}] _{\xa || \yb}
  + \bar S_\omega[\Delta G_{ij}] _{a}
  + 
  \bar S_\omega[\Delta G_{ij}] _{\rm num},
\end{equation}
where 
\begin{align}
  & \bar S_\omega[\Delta G_{ij}] _{\xa || \yb}
    = \normA
    \Big \langle
    \int \limits_{0}^{1}  \frac{{\rm d} \omega}{\omega^{1+2\ep} }
    \left(\psi_{t, \xa j}^{4 \ep} - 1 \right) C_{\xa \yb} 
    \bar S_\omega \left [ \omega^2 \tilde S_{ij}(\xa,\yb) \right ] \Big \rangle_{\xa \yb},
  \\
  &  \bar S_\omega[\Delta G_{ij}] _{a}=
    \frac{\normA}{2}
    \Big \langle
    \int \limits_{0}^{\infty}  \frac{{\rm d} \omega}{\omega^{1-2\ep} }
    f_+(\omega,\psi_{t,\xa j}, \psi_{t,\yb j})
    \frac{(\rho_{\xa \yb} - \rho_{\xa j} - \rho_{\yb j})
    (\rho_{\xa j} - \rho_{\yb j}) }{\rho_{\xa \yb} \rho_{\xa j} \rho_{\yb j}}
    \Big \rangle_{\xa \yb},
    \label{eq5.23}
  \\
  & \bar S_\omega \left [ \Delta G_{ij} \right ]_{\rm num}
    = \frac{\normA}{2}
    \Big \langle
    \int \limits_{0}^{\infty}  \frac{{\rm d} \omega}{\omega}
    \left( \omega^{-2\ep} - \omega^{2\ep} \right)
    f_-(\omega,\psi_{t,\xa j}, \psi_{t,\yb j})
    \bar S_\omega \left [ \omega^2 \tilde S_{ij}(\xa,\yb) \right ] \Big \rangle_{\xa \yb} \\
  & + \normA
    \Big \langle
    \int \limits_{0}^{\infty}  \frac{{\rm d} \omega}{\omega^{1+2\ep} }
    \bar C_{\xa \yb}
    \left[  \left(\psi_{t, \xa j}^{4 \ep} - 1 \right)  \theta( \psi_{t, \xa j }  - \omega \psi_{t,  \yb j}  ) \right] 
    \bar S_\omega \left [ \omega^2 \tilde S_{ij}(\xa,\yb) \right ] \Big \rangle_{\xa \yb}.
    \nonumber 
\end{align}
 
We can compute $\bar S_{\omega}[\Delta G_{ij}]_{\rm num}$ numerically in four
dimensions, whereas $\bar S_{\omega}[\Delta G_{ij}]_{\xa || \yb}$ and $\bar
S_{\omega}[\Delta G_{ij}]_{a}$ need to be discussed further, and we do this in
the follow-up sections.

\subsection{The strongly-ordered term  $S_\omega\left[G_{ij} \right]$}
\label{sec:soLim}

As explained in the previous section, we require the integral of the
strongly-ordered double-emission eikonal function
\begin{equation}
  S_\omega \left [ G_{ij} \right ]
  = \normA
  \left \langle
    \int \limits_{0}^{\infty}
    \frac{{\rm d} \omega}{\omega^{1+2\ep} } \psi_{t, \xa j }^{4 \ep}
    \theta( \psi_{t, \xa j}  - \omega \psi_{t, \yb j}   )
    S_\omega \left [ \omega^2 \tilde S_{ij}(\xa,\yb) \right ] \right \rangle_{\xa \yb}.
\end{equation}
The strongly-ordered eikonal function  evaluates to 
\begin{equation}
  \label{eq:wwSso}
 S_\omega [\omega^2 \tilde S_{ij}(\xa,\yb)]
= \frac{2}{\rho_{\xa \yb} \rho_{\xa j} } + \frac{2}{\rho_{\xa \yb} \rho_{\yb j} } - \frac{2}{\rho_{\xa j} \rho_{\yb j} }
-\frac{1}{\rho_{\xa \yb}} + \frac{1}{\rho_{\yb j}} - \frac{\rho_{\xa j}}{\rho_{\xa \yb} \rho_{\yb j}}.
\end{equation}
Integrating over  $\omega$, we find 
\begin{equation}
  \label{eq:wIntSO}
  \int \limits_{0}^{\infty} \frac{{\rm d} \omega}{\omega^{1+2\ep}}
  \psi_{t,\xa j}^{4 \ep} \theta(\psi_{t, \xa j} - \omega \psi_{t,\yb j}) =
  -\frac{1}{2\ep} \psi_{t, \xa j}^{2\ep} \psi_{t,  \yb j}^{2 \ep}.
\end{equation}
Finally, using the symmetry of the integration measure with respect to the
interchange of $\xa$ and $\yb$, we write the strongly-ordered contribution as
\begin{align}
  \label{eq:wwSsoSym}
  S_\omega\left[G_{ij} \right]
  & = - \frac{\normA}{2\ep}
    \left\langle
    \psi_{t,\xa j }^{2\ep} \psi_{t, \yb j }^{2 \ep}
    \left( 
    \frac{1}{\rho_{\yb j}}
    - \frac{2}{\rho_{\xa j} \rho_{\yb j} }
    -\frac{1}{\rho_{\xa \yb}}
    + \frac{4 - \rho_{ \yb j}}{\rho_{\xa \yb} \rho_{\xa j} }
    \right)
  \right\rangle_{\xa \yb}.
\end{align}
The first two terms in the above expression do not contain $1/\rho_{\xa \yb}$
and can be written using angular integrals defined in
Appendix~\ref{sec:angIntsDef}. We find
\begin{align}
  \label{eq:soInt12}
  \left \langle
  \frac{\psi_{t, \xa j }^{2\ep} \psi_{t, \yb  j }^{2 \ep}}{\rho_{\yb j}}
  \right \rangle_{\xa \yb}
  & =  \rho_{tj}^{-4\ep} \AIm{-2\ep}{}\AImz{-2\ep}{1},\\
  \left \langle
  \frac{\psi_{t, \xa j}^{2\ep} \psi_{t, \yb j}^{2 \ep}}{\rho_{\xa j}\rho_{\yb j}}
  \right \rangle_{\xa \yb}
  & =  \rho_{tj}^{-4\ep} \left( \AImz{-2\ep}{1} \right)^2.
\end{align}
We note that we do not show arguments of the above integrals because they are
always the same, i.e. $I^{(1)}_{a,b} = I^{(1)}_{a,b}\left[\rho_{tt},\rho_{tj}\right]$ and
$I^{(1)}_a = I^{(1)}_a\left[\rho_{tt}\right] $, where $\rho_{tt} = 1-\beta^2$ and $\rho_{tj} = (1-\beta^2)/(1-\beta \csth)$.

Using the $\xa \leftrightarrow \yb$ symmetry, we write the third term in
Eq.~(\ref{eq:wwSsoSym}) as
\begin{equation}
  \begin{split}
    \label{eq:soInt3}
    \left \langle
    \frac{\psi_{t, \xa j }^{2\ep} \psi_{t,\yb j}^{2 \ep}}{\rho_{\xa \yb}}
    \right \rangle_{\xa \yb}
    & =
      \left \langle
      \frac{\psi_{t, \xa j }^{4\ep}}{\rho_{\xa \yb}}
      \right \rangle_{\xa \yb}
      - \frac{1}{2} \left \langle
      \left( \psi_{t, \xa j}^{2\ep} - \psi_{t,  \yb j}^{2 \ep} \right)^2
      \frac{1}{\rho_{\xa \yb}}
      \right \rangle_{\xa \yb}
    \\
    & = \rho_{tj}^{-4\ep}  \AIm{-4\ep}
      - \frac{\ep^2}{2} \left \langle
      \frac{ \left( g^{(2)}_{t, \xa j} - g^{(2)}_{t,  \yb j} \right)^2}{\rho_{\xa \yb}}
      \right \rangle_{\xa \yb},
  \end{split}
\end{equation}
where in the last step we have used $\psi_{t, \xa j}^{2\ep} = 1+ \ep g_{t, \xa j
}^{(2)}$. We emphasize that this equation is the definition of $g^{(2)}_{t, \xa
j}$, and not the expansion of $\psi_{t,\xa j}^{2\ep}$ to first power in $\ep$.
Finally, we note that the last term in Eq.~(\ref{eq:soInt3}) gives a finite
contribution to the final result for the integrated double-emission eikonal
function, and can be numerically integrated in four dimensions.

The last term in~Eq.~\eqref{eq:wwSsoSym} is more complicated. To simplify it, we
first substitute $\psi_{t, \xa j}^{2\ep} = 1+ \ep g_{t,\xa j }^{(2)}$, $\psi_{t,
\yb j}^{2\ep} = 1+ \ep g_{t, \yb j }^{(2)}$ there, and make use of the fact
that, unless two functions $g^{(2)}$ appear in the integrand, integration over
$\xa$ or $\yb$ can be performed. Hence, we write
\begin{align}
  \label{eq:soInt4}
  \left \langle
  \psi_{t, \xa j }^{2\ep} \psi_{t, \yb j }^{2 \ep} \frac{4-\rho_{\yb j}}{\rho_{\xa \yb}\rho_{\xa j}}
  \right \rangle_{\xa \yb}
  & =   T_{4,0} +   T_{4,1} +   T_{4,2},
\end{align}
where  
\begin{align}
  \label{eq:soInt4split}
  T_{4,0} &=\frac{1}{2\ep} \rho_{tj}^{-2\ep} \left(
        \AIm{-2\ep}- 2( 2 - 3\ep)\AImz{-2\ep}{1}
        \right),
  \\
  T_{4,1} & = \ep \left \langle
        g_{t,\xa j}^{(2)} (4 - \rho_{\xa j}) \AIzz{1}{1}{\rho_{\xa j}}
        \right \rangle_{\xa},
  \\
  T_{4,2} & =  \ep^2 \left \langle
        g_{t, \xa j }^{(2)} g_{t,\yb j }^{(2)}  \frac{4-\rho_{\yb j}}{\rho_{\xa \yb}\rho_{\xa j}}
        \right \rangle_{\xa \yb}.
\end{align}
Using the integral $\AIzz{1}{1}{\rho_{\xa j}}$ from Eq.~\eqref{eq:angInt-2-00},
and transformation rules for hypergeometric functions, the integral $T_{4,1}$
can be cast into the following form
\begin{equation}
  \label{eq:soInt41}
  T_{4,1} = 2^\ep(1-2\ep)\left \langle g_{t,\xa j}^{(2)} \frac{(\rho_{\xa j} - 4)}{\rho_{\xa j}^{1+\ep}}{}_2F_1\left(-\ep,-\ep;1-\ep; 1 - \frac{\rho_{\xa j}}{2} \right) \right \rangle_{\xa}.
\end{equation}
Expanding in $\ep$ to the required order, it is possible to replace the
hypergeometric function with the following expression
\begin{equation}
  \label{eq:h21z2zExp}
{}_2F_1\left(-\ep,-\ep;1-\ep; 1 - \frac{\rho_{\xa j}}{2} \right) = 1 + \ep^2 \mathrm{Li}_2 \left(1 - \frac{\rho_{\xa j}}{2} \right) + \mathcal{O}\left( \ep^3 \right).
\end{equation}
The final result for the integral $T_{4,1}$ valid through ${\cal O}(\ep^2)$ then
easily follows. It reads
\begin{equation}
  \begin{split}
    \label{eq:soInt41exp}
    T_{4,1} = & - \frac{2^\ep (1-2\ep)^2}{\ep^2}\AIz{\ep} 
                +\frac{2^\ep (1-2\ep)}{\ep} \rho_{t j}^{-2\ep} \left(\AImz{-2\ep}{\ep} -4 \AImz{-2\ep}{1+\ep} \right)\\
              & -2\ep^2 \left \langle  \frac{4-\rho_{\xa j}}{\rho_{\xa j}}
                {\rm Li}_2 \left (1 - \frac{\rho_{\xa j}}{2}
                \right ) \log \frac{\rho_{t \xa }}{\rho_{t j}} 
                \right \rangle_\xa.
  \end{split}
\end{equation}

The integral $T_{4,2}$ diverges for $\xa ||\yb$; hence, we subtract this limit
by inserting $1 = C_{\xa \yb} + (1-C_{\xa \yb})$ into the integrand. Since
\begin{equation}
  \label{eq:soI4cmnAction}
  C_{\xa \yb} \; g^{(2)}_{t,\xa j} g^{(2)}_{t,\yb j} \frac{(4 - \rho_{\yb j})  }{\rho_{\xa \yb} \rho_{\xa j}} =
  \left ( g^{(2)}_{t,\xa j} \right )^2\frac{(4-\rho_{\xa j})}{\rho_{\xa \yb} \rho_{\xa j}},
\end{equation}
we find 
\begin{align}
  \label{eq:soI4cmnIntegrated}
  \Big \langle & C_{\xa \yb} \; \ep^2 \; g^{(2)}_{t,\xa j} g^{(2)}_{t,\yb j} \frac{(4 - \rho_{\yb j})  }{\rho_{\xa \yb} \rho_{\xa j}} \Big \rangle_{\xa \yb}
   = \frac{(1-2\ep)(2-3\ep)}{2\ep^2} 
  \nonumber\\
  & + \frac{1-2\ep}{2\ep} 
    \rho_{tj}^{-4\ep}\left(
    \AIm{-4\ep} - 4\AImz{-4\ep}{1}
    \right)
 - \frac{1-2\ep}{\ep} 
    \rho_{tj}^{-2\ep}\left(
    \AIm{-2\ep}- 4\AImz{-2\ep}{1}
    \right).
\end{align}
The remaining integral with $\bar C_{\xa \yb} = 1 - C_{\xa \yb}$ becomes finite,
and can be calculated in four dimensions numerically. We obtain
\begin{equation}
   \label{eq:soInt4cmnBar}
   \left \langle {\overline C}_{\xa \yb} \; \ep^2 \; g^{(2)}_{t,\xa j} g^{(2)}_{t,\yb j} \frac{(4 - \rho_{\yb j})}{\rho_{\xa \yb} \rho_{\xa j}} \right \rangle_{\xa \yb} =
 \ep^2   \left \langle  g^{(2)}_{t,\xa j}  \frac{g^{(2)}_{t,\yb j}(4 - \rho_{\yb j}) - g^{(2)}_{t,\xa j}(4 - \rho_{\xa j})}{\rho_{\xa \yb} \rho_{\xa j}} \right \rangle_{\xa \yb}.
\end{equation}

The combined result for the strongly-ordered limit reads
\begin{equation}
  \begin{split}
    & \label{eq:soLimComb}
      \frac{S_\omega\left[G_{ij} \right]}{\normA}
      =
      -\frac{(1-2 \ep) (2-3 \ep)}{4 \ep^3}
      + \frac{(1-2 \ep)^2}{2^{1-\ep} \ep^3}\AIz{\ep}
      + \frac{1}{2\ep} \rho_{tj}^{-4\ep} \AImz{-2\ep}{1} \left(2 \AImz{-2\ep}{1} - \AIm{-2\ep}  \right)
    \\
    & + \frac{(1-2\ep)}{2\ep^2} \rho_{tj}^{-4\ep} \left(2 \AImz{-4\ep}{1} - \AIm{-4\ep}  \right)
      + \frac{(1-2\ep)}{2^{1-\ep} \ep^2} \rho_{tj}^{-2\ep} \left(4 \AImz{-2\ep}{1+\ep} - \AImz{-2\ep}{\ep}  \right)
    \\
    & + \frac{1}{4 \ep^2} \rho_{tj}^{-2\ep} \left( (1-4\ep) \AIm{-2\ep} - 2(2-5\ep) \AImz{-2\ep}{1}  \right)
      +\ep 
      \left \langle 
      \frac{4-\rho_{\xa j}}{\rho_{\xa j}}
      {\rm Li}_2 \left (1 - \frac{\rho_{\xa j}}{2}
      \right ) \log \frac{\rho_{t \xa }}{\rho_{t j}} 
      \right \rangle_\xa
    \\
    & 
      -\frac{\ep}{4} \Big \langle
      \frac{(g^{(2)}_{t,\xa j} - g^{(2)}_{t,\yb j})^2}{\rho_{\xa\yb}}
      \Big \rangle_{\xa \yb}   
      -\frac{\ep}{2} \Big \langle
      g^{(2)}_{t,\xa j} \frac{
      g^{(2)}_{t,\xa j} (\rho_{\xa j} -4)
      - g^{(2)}_{t,\yb j} (\rho_{\yb j}- 4)
      }{\rho_{\xa \yb} \rho_{\xa j}}
      \Big \rangle_{\xa \yb}.
  \end{split}
\end{equation}

\subsection{Calculation of 
$\bar{S}_\omega \left[ G_{ij}^{(0)}\right]$ }
\label{sec:intG0}

It is straightforward to compute the term $\bar{S}_\omega \left[ G_{ij}^{(0)}
\right]$ in Eq.~(\ref{eq5.20}), which does not depend on the velocity $\beta$
and the angle between $\vec n_i$ and $\vec n_j$. It can be expressed in terms of
the integrals discussed in Section~\ref{sec:direct-calculation-P-rest} if we
choose $p_i = (m, \vec 0)$ in that section. We obtain
\begin{equation}
  \begin{split}
    \label{eq:wwSbar0red}
    & \frac{\bar{S}_\omega \left[ G_{ij}^{(0)} \right]}{\normA} 
      =  -\frac{11}{24 \ep^2}
      +\frac{1}{\ep}\biggl(
      \frac{65}{36}
      -\frac{11}{6}\log(2)
      -\frac{\pi^2}{24}
      \biggr)
      -\biggl(
      \frac{553}{108}
      -\frac{269}{36}\log(2)
      -\frac{\pi^2}{12} 
    \\
    & 
      -\frac{11}{6} \log^2(2)
      +\frac{11}{8} \zeta_3
      \biggr)
    +\ep \biggl(
    \frac{3887}{324}
    -\frac{1133}{54}\log(2)
    -\frac{47}{144}\pi^2
    -\frac{341}{36}\log^2(2) \\
  &
    -\frac{11}{9}\log^3(2)
    +\frac{11}{36} \pi^2 \log(2)
     +\frac{99}{8} \zeta_3
    -\frac{29}{1440}\pi^4
    +\frac{1}{12} \pi^2 \log^2(2)
    -\frac{1}{12} \log^4(2) \\
  &
    -\frac{7}{4} \zeta_3 \log(2)
    -2 \text{Li}_4\left(\frac{1}{2}\right)
    \biggr)
    +\mathcal{O}\left(\ep^2\right).
  \end{split}
\end{equation}

\subsection{Integration of the double-collinear subtraction term}
\label{sec:dcSubtr}
We proceed with the calculation of the integral of the 
double-collinear $\xa || \yb$ subtraction term. It reads 
\begin{equation}
  \label{eq:dcIntSectors}
  \bar S_\omega[\Delta G_{ij}] _{\xa || \yb}=
  {\cal N}_A\; \Big \langle
  \int \limits_{0}^{1}  \frac{{\rm d} \omega}{\omega^{1+2\ep} }
  \left(\psi_{t, \xa j}^{4 \ep} - 1 \right) C_{\xa \yb}  
  \bar S_\omega \left [ \omega^2 \tilde S_{ij}(\xa,\yb) \right ] \Big \rangle_{\xa \yb}.
\end{equation}
The soft-subtracted eikonal function $\bar S_\omega [\omega^2 \tilde
S_{ij}(\xa,\yb)]$ has a singularity in the limit $\xa||\yb$. Isolating this
singularity requires care as it appears, naively, as the second-order pole in
$\rho_{\xa \yb}$ whereas the pole is, in fact, first-order.

The relevant limit can be easily computed. We obtain 
\begin{align}
  C_{\xa \yb} \bar S_\omega \left [ \omega^2 \tilde S_{ij}(\xa,\yb) \right ] =
  \frac{4(1-\ep)\omega^2 (\vec n_j \cdot \vec \kappa)^2}{(1+\omega)^4  \rho_{\xa\yb} \rho_{\xa j}^2}
  - \frac{4\omega (2 - \rho_{\xa j})}{(1+\omega)^2 \rho_{\xa \yb}\rho_{\xa j}},
\end{align}
where  $\vec \kappa$ is defined by the following equation 
\begin{equation}
  \label{eq:kPrepDef}
  \vec{n}_\yb = \cos{\theta_{\xa \yb}} \vec{n}_\xa + \sin{\theta_{\xa \yb}} \; \vec \kappa,
\;\;  \;\; \vec \kappa \cdot \vec n_\xa = 0,
  \;\;\; \vec \kappa^2 = 1.
\end{equation}

To integrate over directions of parton $\yb$ in Eq.~(\ref{eq:dcIntSectors}), we
first average over directions of $\vec{\kappa}$. Using
\begin{equation}
  \kappa^i \kappa^j \to 
  \frac{\delta^{ij} -\vec n^i_\xa \vec n^j_\xa}{d-2},
\end{equation}
we obtain 
\begin{align}  
  \left \langle C_{\xa \yb} \bar S_\omega \left [ \omega^2 \tilde S_{ij}(\xa,\yb) \right ] 
  \right \rangle_{\kappa}
  =
  -2\frac{(2-\rho_{\xa j}) }{\rho_{\xa j} \rho_{\xa \yb} }
  \;
  \frac{ \omega  (2 + 3 \omega + 2 \omega^2)
  }{(1+\omega)^4}.
\end{align}

Since integrations over energies and angles factorize, it is straightforward to
complete the calculation. We find
\begin{equation}
  \begin{split}
    \label{eq:dcIntResults}
    &  
      \frac{ S_\omega[\Delta G_{ij}] _{\xa || \yb}
      }{{\cal N}_A}
      =
      \frac{(1-\ep)(1-2\ep)}{\ep^2} \gamma_\omega
      + \frac{1-2\ep}{\ep} \gamma_\omega 
      \rho_{tj}^{-4\ep}\left(
      2\AImz{-4\ep}{1}
      - \AIm{-4\ep} 
      \right),
  \end{split}
\end{equation}
where 
\begin{equation}
  \begin{split}
    \gamma_\omega  =&  \int \limits_{0}^{1}  \frac{{\rm d} \omega}{\omega^{1+2\ep} }
                      \frac{\omega(2+3\omega+2\omega^2)}{(1+\omega)^4} =
                      \frac{11}{12}
                      + \ep \left(\frac{1}{12}+\frac{11}{3}\log(2)\right)\\
                    & 
                      + \ep^2 \left(\frac{1}{3}+\frac{11}{18}\pi^2\right) 
                      + \ep^3 \left(\frac{4}{3}\log(2) + 11 \zeta_3 \right)
                      +\mathcal{O}\left(\ep^4\right).
  \end{split}
\end{equation}

\subsection{Calculation  
of $\bar S_\omega[\Delta G_{ij}]_{a}$}
\label{sec:ga}

To compute $\bar S_\omega[\Delta G_{ij}]_{a}$ defined in Eq.~(\ref{eq5.23}), we
integrate over $\omega$ and combine the obtained result with the one where $\xa$
and $\yb$ are interchanged. We find
\begin{equation}
  \label{eq:SwGa}
  \bar S_\omega[\Delta G_{ij}]_{a} = \frac{\normA}{8\ep}
  \left \langle 
    \left ( 
      \left ( \frac{\psi_{t,\yb j}}{\psi_{t,\xa j}} 
      \right )^{2\ep}
      - \left ( \frac{\psi_{t,\xa j}}{\psi_{t,\yb j}} 
      \right )^{2\ep}
    \right )
    \frac{(\rho_{\xa \yb} - \rho_{\xa j} - \rho_{\yb j})
      (\rho_{\xa j} - \rho_{\yb j}
      ) }{\rho_{\xa \yb} \rho_{\xa j} \rho_{\yb j}}
  \right \rangle_{\xa\yb}.
\end{equation}
To compute this integral, we expand in $\ep$ through the relevant order
\begin{equation}
  \frac{1}{8\ep}
  \left ( 
    \left ( \frac{\psi_{\yb j}}{\psi_{\xa j}} 
    \right )^{2\ep}
    - \left ( \frac{\psi_{\xa j}}{\psi_{\yb j}} 
    \right )^{2\ep}
  \right )  
  = \frac{1}{4} 
  \left ( g_{t,\yb j}^{(2)}
    - g_{t,\xa  j}^{(2)} \right )
  -\frac{\ep}{8} 
  \left ( \left ( g_{t,\yb j}^{(2)} \right )^2
    - \left ( g_{t,\xa  j}^{(2)} \right )^2 \right ) +{\cal O}(\ep^2),
\end{equation}
and get rid of $g^{(2)}_{t,\xa j}$ terms by using the $\xa \leftrightarrow \yb$
permutation symmetry of the integrand in Eq.~(\ref{eq:SwGa}). 
We find
\begin{equation}
  \bar S_\omega[\Delta G_{ij}]_a =  \frac{\normA}{2}
  \left \langle 
    \left ( 
      g_{t,\yb j}^{(2)} - \frac{\ep}{2} \left ( g_{t,\yb j}^{(2)} \right )^2
    \right ) 
    \frac{(\rho_{\xa \yb} - \rho_{\xa j} - \rho_{\yb j})
      (\rho_{\xa j} - \rho_{\yb j}
      ) }{\rho_{\xa \yb} \rho_{\xa j} \rho_{\yb j}}
  \right \rangle_{\xa\yb}.
\end{equation}

Integration over $\xa$ 
can  be easily performed. We obtain
\begin{equation}
  \left \langle \frac{(\rho_{\xa \yb} - \rho_{\xa j} - \rho_{\yb j})
      (\rho_{\xa j} - \rho_{\yb j}
      ) }{\rho_{\xa \yb} \rho_{\xa j} \rho_{\yb j}}
  \right 
  \rangle_{\xa} 
  = \frac{1-\ep}{\ep}  + \rho_{\yb j} I_{11}^{(0)}(\rho_{\yb j}).
\end{equation}
 
Using 
\begin{equation}
  g_{t,\yb j}^{(2)} - \frac{\ep}{2} \left ( g_{t,\yb j}^{(2)} \right )^2 = 
  \frac{1}{2\ep} 
  \left (\psi_{t,\yb j}^{2\ep} - \psi_{t,\yb j}^{-2\ep}
  \right )
  +{\cal O}(\ep^2),
\end{equation}
we find
\begin{equation}
  \bar S_\omega[\Delta G_{ij}]_a = 
  \frac{\normA}{4 \ep}
  \left \langle 
    \left (\psi_{t,\yb j}^{2\ep} - \psi_{t,\yb j}^{-2\ep}
    \right )
    \left ( 
      \frac{1-\ep}{\ep}  + \rho_{\yb j} I_{11}^{(0)}(\rho_{\yb j}) 
    \right )
  \right \rangle_\yb. 
\end{equation}
Performing the remaining integrations, we obtain 
\begin{equation}
  \begin{split}
    \frac{ \bar S_\omega[\Delta G_{ij}]_a
    }{\normA}
    & = \frac{1-\ep}{4\ep^2} 
      \left(
      \rho_{tj}^{-2\ep}\AIm{-2\ep} - \rho_{tj}^{2\ep}\AIm{2\ep}
      \right)
    \\
    & -\frac{1-2\ep}{8\ep^2}
      \left \langle \left ( 
      \rho_{tj}^{-2\ep}
      \rho_{t,\yb }^{2\ep} 
      -\rho_{tj}^{2\ep}
      \rho_{t,\yb }^{-2\ep} 
      \right )
      \;  \rho_{\yb j} \; _2F_1(1,1,1-\ep,\frac{\rho_{\yb \bar j}}{2} )
      \right \rangle_{\yb} 
      \label{eq5.57}.
  \end{split}
\end{equation}
We use transformation properties of the hypergeometric function to rewrite the above result as follows
\begin{equation}
  \begin{split}
    & \frac{ \bar  S_\omega[\Delta G_{ij}]_a
      }{\normA} 
      = \frac{1-\ep}{4\ep^2} 
      \left(
      \rho_{tj}^{-2\ep}\AIm{-2\ep}  - \rho_{tj}^{2\ep}\AIm{2\ep}
      \right)
    \\
    & -\frac{1-2\ep}{2^{2-\ep} \ep^2}
      \left ( 
      \rho_{tj}^{-2\ep}
      I^{(1)}_{-2\ep,\ep}
      - 
      \rho_{tj}^{2\ep}
      I^{(1)}_{2\ep,\ep}
      \right ) 
      -\ep \left \langle 
      \log \frac{\rho_{t,\yb}}{\rho_{t j}}
      \; {\rm Li}_2\left (1-
      \frac{\rho_{\yb  j}}{2} \right ) 
      \right \rangle_{\yb} + {\cal O}(\ep^2). 
  \end{split}
  \label{eq5.50}
\end{equation}

\subsection{Combined expression for the double-soft integral}

We are now in a position to present the result for the integrated
double-emission eikonal function where the divergent part is extracted
analytically, and the finite part is written as an integral that can be computed
numerically in straightforward way. Our starting point is
Eq.~\eqref{eq:boostFrameSS} which we repeat here
\begin{equation}
  \intSS[{\tilde S}_{ij}]
  = G_{ij} = 
  S_\omega[G_{ij}] + \bar S_\omega [ G_{ij}^{(0)}]
  + \bar S_\omega[\Delta G_{ij}].
\end{equation}
The result for $S_\omega [G_{ij}]$ can be found in Eq.~(\ref{eq:soLimComb}) and
the result for $\bar S_\omega [G^{(0)}_{ij}]$ in Eq.~(\ref{eq:wwSbar0red}). The
quantity $\bar S_\omega[\Delta G_{ij}]$ is defined in
Eq.~(\ref{eq:barSwDeltaG}). There are three terms there. The result for $\bar
S_\omega[\Delta G_{ij}]_{\xa || \yb} $ is given in Eq.~(\ref{eq:dcIntResults}),
the result for $\bar S_\omega[\Delta G_{ij}]_{a}$ can be found in
Eq.~(\ref{eq5.57}), and $\bar S_\omega[\Delta G_{ij}]_{\rm num}$ is finite and
can be computed numerically.

To present the final result, we separate contributions that are known
analytically $(R)$ and the ones that are finite and can be integrated
numerically $(N)$. We write
\begin{equation}
  \label{eq:SSfinal-pA}
  \intSS\left[ \tilde{S}_{ij} \right]  = \normA \left( R + N \right),
\end{equation}
where  
\begin{equation}
  \begin{split}
    R = & \frac{\bar{S}_\omega \left[ G_{ij}^{(0)} \right]}{\normA} -\frac{(1 - 2 \ep) (2 - 3 \ep)}{
          4 \ep^3} + \frac{(1 - \ep) (1 - 2 \ep) \gamma_\omega}{\ep^2}
          + \frac{ (1 - 2 \ep)^2}{2^{1-\ep}\ep^3} I_\ep^{(0)}
    \\
        & -\frac{(1-2\ep)}{2\ep^2} \rho_{tj}^{-4\ep}
          (1+2\ep \gamma_\omega)
          I^{(1)}_{-4\ep}
          -\frac{1-\ep}{4\ep^2} \rho_{tj}^{2\ep} I^{(1)}_{2\ep}
          +\frac{(1-2\ep)}{\ep^2} \rho_{tj}^{-4\ep}
          (1+2\ep \gamma_\omega)
          I^{(1)}_{-4\ep,1}
    \\
        &  -\frac{2-5\ep}{2\ep^2} \rho_{tj}^{-2\ep}I^{(1)}_{-2\ep,1}
          +\frac{1}{\ep}\rho_{tj}^{-4\ep}\left ( I^{(1)}_{-2\ep,1} \right )^2
          +\frac{(2-5\ep)}{4\ep^2} \rho_{tj}^{-2\ep}I^{(1)}_{-2\ep}
          -\frac{1}{2\ep} \rho_{tj}^{-4\ep} 
          I^{(1)}_{-2\ep} I^{(1)}_{-2\ep,1}
    \\
        & -\frac{3(1-2\ep)}{2^{2-\ep} \ep^2}\rho_{tj}^{-2\ep} I^{(1)}_{-2\ep,\ep}
          +\frac{(1-2\ep)}{2^{2-\ep} \ep^2}
          \left ( 8 \rho_{tj}^{-2\ep} I^{(1)}_{-2\ep,1+\ep}
          + \rho_{tj}^{2\ep} I^{(1)}_{2\ep,\ep}
          \right ), 
  \end{split}
\end{equation}
and 
\begin{align}
  N & = -\ep  \left \langle
      \frac{
      \log^2\rho_{t \xa}/\rho_{t \yb} }{\rho_{\xa \yb}}
      +\frac{2 \log{\psi_{\xa j}} \left(
      \log{\psi_{\xa j}}(\rho_{\xa j} - 4) - \log{\psi_{\yb j}}(\rho_{\yb j} - 4)  \right) }{\rho_{\xa \yb} \rho_{\xa j}}
      \right \rangle_{\xa \yb}
      \nonumber \\
    &  -2 \ep  
      \Big \langle
      \int \limits_{0}^{\infty}  \frac{{\rm d} \omega \log{\omega}}{\omega}
      \left(\theta( \psi_{t, \xa j }  - \omega \psi_{t,  \yb j}  ) - \theta(1-\omega)\right) 
      \bar S_\omega \left [ \omega^2 \tilde S_{ij}(\xa,\yb) \right ]
      \Big \rangle_{\xa \yb}
      \label{eq6.4}     \\
    & + 4 \ep 
      \Big \langle
      \int \limits_{0}^{\infty}  \frac{{\rm d} \omega \log{\psi_{t, \xa j}}}{\omega}
      \left(
      \theta( \psi_{t, \xa j }  - \omega \psi_{t,  \yb j}  )
      - \theta(1-\omega)
      C_{\xa \yb} 
      \right)\bar S_\omega \left [ \omega^2 \tilde S_{ij}(\xa,\yb) \right ]
      \Big \rangle_{\xa \yb}
      \nonumber\\
    & + \ep 
      \left \langle 
      \frac{4 - 2 \rho_{\xa j}}{\rho_{\xa j}}
      {\rm Li}_2 \left ( 1- \frac{\rho_{\xa j}}{2}
      \right )
      \log \frac{\rho_{t \xa}}{\rho_{t j}}
      \right \rangle_\xa. 
      \nonumber 
\end{align}

We have used the above representation to compute the integrated double-emission
eikonal function $\intSS[\tilde S_{ij}]$ for various values of $\beta$ and $\cos
\theta$ and found excellent agreement with the results reported in
Section~\ref{sec:direct-calculation-P-rest}. We emphasize that all divergent
contributions are known analytically\footnote{Explicit results for integrals
$I^{(1)}_{\alpha, \beta}$ can be found in Appendix~\ref{sec:angIntsDef} and in
the ancillary file. } and that only integrals in Eq.~(\ref{eq6.4}) need to be
evaluated numerically. The quality and the speed of this numerical integration
are acceptable; obviously, it depends rather strongly on values of $\beta$ and
$\cos \theta$, with values $\beta \sim 1$ being the most challenging.

\section{Conclusion}
\label{sec:concl}
We have described the computation of the integrated double-emission eikonal
function with one massive and one massless emitter. This quantity is a required
ingredient for extending the nested soft-collinear subtraction scheme
\cite{Caola:2017dug} to processes with massive particles. We have shown that the
extension of reverse unitarity to cases with Heaviside functions in the
integrand \cite{Baranowski:2021gxe} allows one to simplify the calculation
significantly, and derive the differential equations for relevant master
integrals in a straightforward way. Although the resulting differential
equations contain two sectors with elliptic integrals, we were able to remove
them from the final result and express it in terms of ordinary polylogarithms
and ${\rm Li}_{2,2}$. As such, the result appears to be structurally similar to
the one reported in Ref.~\cite{Caola:2018pxp}, but it is significantly more
complex. Since, even after all the simplifications of the final result, it
remains complicated, we have developed fast and efficient \texttt{C} code that can be
used to compute the result for an arbitrary kinematic point with high precision.

Finally, we have studied the possibility to calculate the same quantity in an
entirely different way, partially motivated by the calculation of the
$N$-jettiness soft function in Ref.~\cite{Agarwal:2024gws}. The idea is to
perform the computation in the rest frame of a massive parton, where the
dependence on the original angle between two partons appears in kinematic
constraints only. Identifying singularities of the double-emission eikonal
function in this frame, allows us to compute the divergent contributions with
relative ease, and design a simple representation for the finite remainder that
can be integrated numerically right away. Apart from providing an important
cross check for the calculation based on reverse unitarity and differential
equations, this approach can open the way for computing integrated subtracted
terms independently of the complexity of an ``observable'', used to define them.
We look forward to further applications of this methodology in the future.

\section*{Acknowledgments}
 D.H. thanks Institute for Theoretical Particle Physics (TTP) at KIT  for support during the early stages of this project.
 K.M.  is grateful to  CERN Theory Department  for hospitality extended to him during work on this paper.
This research was supported in part by  
Deutsche Forschungsgemeinschaft (DFG, German Research Foundation) under grant no.\ 396021762 - TRR 257.

\appendix

\section{Master integrals}
\label{sec:MIs}
In this appendix, the complete list of master integrals  is given. 

\begin{alignat}{3}
  \label{eq:dtMIs}
  & J_{1} && = \dInt {0,0,0}{0,0,0}{0}
  && = \int [\dm l_{\xa}] [\dm l_{\yb}]\delta\left(1 - l_{\xa}\cdot P\right)\delta\left(1 - l_{\yb}\cdot P  \right),
  \\
  & J_{2} && = \dInt {0,0,0}{0,0,1}{0}
  && = \int [\dm l_{\xa}] [\dm l_{\yb}] \frac{\delta\left(1 - l_{\xa}\cdot P\right)\delta\left(1 - l_{\yb}\cdot P  \right)}{(l_{\xa\yb}\cdot p_j)},
  \\
  & J_{3} && = \dInt {0,0,1}{0,0,0}{0}
  && = \int [\dm l_{\xa}] [\dm l_{\yb}] \frac{\delta\left(1 - l_{\xa}\cdot P\right)\delta\left(1 - l_{\yb}\cdot P  \right)}{(l_{\xa\yb}\cdot p_i)},
 \\
  & J_{4} && = \dInt {0,0,2}{0,0,0}{0}
  && = \int [\dm l_{\xa}] [\dm l_{\yb}] \frac{\delta\left(1 - l_{\xa}\cdot P\right)\delta\left(1 - l_{\yb}\cdot P  \right)}{(l_{\xa\yb}\cdot p_i)^2},
 \\
  & J_{5} && = \dInt {1,0,0}{0,0,0}{0}
  && = \int [\dm l_{\xa}] [\dm l_{\yb}] \frac{\delta\left(1 - l_{\xa}\cdot P\right)\delta\left(1 - l_{\yb}\cdot P  \right)}{(l_{\xa}\cdot p_i)},
 \\
  & J_{6} && = \dInt {1,0,0}{0,1,0}{1}
  && = \int [\dm l_{\xa}] [\dm l_{\yb}] \frac{\delta\left(1 - l_{\xa}\cdot P\right)\delta\left(1 - l_{\yb}\cdot P  \right)}{(l_{\xa}\cdot p_i)(l_{\yb}\cdot p_j)(l_{\xa}\cdot l_{\yb})},
 \\
  & J_{7} && = \dInt {1,0,0}{0,2,0}{1}
  && = \int [\dm l_{\xa}] [\dm l_{\yb}] \frac{\delta\left(1 - l_{\xa}\cdot P\right)\delta\left(1 - l_{\yb}\cdot P  \right)}{(l_{\xa}\cdot p_i)(l_{\yb}\cdot p_j)^2(l_{\xa}\cdot l_{\yb})},
 \\
  & J_{8} && = \dInt {2,0,0}{0,1,0}{1}
  && = \int [\dm l_{\xa}] [\dm l_{\yb}] \frac{\delta\left(1 - l_{\xa}\cdot P\right)\delta\left(1 - l_{\yb}\cdot P  \right)}{(l_{\xa}\cdot p_i)^2(l_{\yb}\cdot p_j)(l_{\xa}\cdot l_{\yb})},
 \\
  & J_{9} && = \dInt {1,0,0}{1,0,0}{0}
  && = \int [\dm l_{\xa}] [\dm l_{\yb}] \frac{\delta\left(1 - l_{\xa}\cdot P\right)\delta\left(1 - l_{\yb}\cdot P  \right)}{(l_{\xa}\cdot p_i)(l_{\xa}\cdot p_j)},
 \\
  & J_{10} && = \dInt {0,0,1}{0,1,0}{0}
  && = \int [\dm l_{\xa}] [\dm l_{\yb}] \frac{\delta\left(1 - l_{\xa}\cdot P\right)\delta\left(1 - l_{\yb}\cdot P  \right)}{(l_{\xa\yb}\cdot p_i)(l_{\yb}\cdot p_j)},
 \\
  & J_{11} && = \dInt {0,0,1}{0,1,-1}{0}
  && = \int [\dm l_{\xa}] [\dm l_{\yb}] \frac{\delta\left(1 - l_{\xa}\cdot P\right)\delta\left(1 - l_{\yb}\cdot P  \right)(l_{\xa\yb}\cdot p_j)}{(l_{\xa\yb}\cdot p_i)(l_{\yb}\cdot p_j)},
 \\
  & J_{12} && = \dInt {0,0,1}{0,1,0}{1}
  && = \int [\dm l_{\xa}] [\dm l_{\yb}] \frac{\delta\left(1 - l_{\xa}\cdot P\right)\delta\left(1 - l_{\yb}\cdot P  \right)}{(l_{\xa\yb}\cdot p_i)(l_{\yb}\cdot p_j)(l_{\xa}\cdot l_{\yb})},
 \\
  & J_{13} && = \dInt {0,0,1}{0,2,0}{1}
  && = \int [\dm l_{\xa}] [\dm l_{\yb}] \frac{\delta\left(1 - l_{\xa}\cdot P\right)\delta\left(1 - l_{\yb}\cdot P  \right)}{(l_{\xa\yb}\cdot p_i)(l_{\yb}\cdot p_j)^2(l_{\xa}\cdot l_{\yb})},
 \\
  & J_{14} && = \dInt {0,0,2}{0,1,0}{1}
  && = \int [\dm l_{\xa}] [\dm l_{\yb}] \frac{\delta\left(1 - l_{\xa}\cdot P\right)\delta\left(1 - l_{\yb}\cdot P  \right)}{(l_{\xa\yb}\cdot p_i)^2(l_{\yb}\cdot p_j)(l_{\xa}\cdot l_{\yb})},
 \\
  & J_{15} && = \dInt {0,0,1}{1,1,0}{0}
  && = \int [\dm l_{\xa}] [\dm l_{\yb}] \frac{\delta\left(1 - l_{\xa}\cdot P\right)\delta\left(1 - l_{\yb}\cdot P  \right)}{(l_{\xa\yb}\cdot p_i)(l_{\xa}\cdot p_j)(l_{\yb}\cdot p_j)},
 \\
  & J_{16} && = \dInt {1,0,1}{0,0,0}{0}
  && = \int [\dm l_{\xa}] [\dm l_{\yb}] \frac{\delta\left(1 - l_{\xa}\cdot P\right)\delta\left(1 - l_{\yb}\cdot P  \right)}{(l_{\xa}\cdot p_i)(l_{\xa\yb}\cdot p_i)},
 \\
  & J_{17} && = \dInt {1,0,1}{0,1,0}{0}
  && =\int [\dm l_{\xa}] [\dm l_{\yb}] \frac{\delta\left(1 - l_{\xa}\cdot P\right)\delta\left(1 - l_{\yb}\cdot P  \right)}{(l_{\xa}\cdot p_i)(l_{\xa\yb}\cdot p_i)(l_{\yb}\cdot p_j)},
 \\
  & J_{18} && = \dInt {1,1,0}{0,0,0}{0}
  && = \int [\dm l_{\xa}] [\dm l_{\yb}] \frac{\delta\left(1 - l_{\xa}\cdot P\right)\delta\left(1 - l_{\yb}\cdot P  \right)}{(l_{\xa}\cdot p_i)(l_{\yb}\cdot p_i)},
 \\
  & J_{19} && = \dInt {1,1,0}{1,0,0}{0}
  && = \int [\dm l_{\xa}] [\dm l_{\yb}] \frac{\delta\left(1 - l_{\xa}\cdot P\right)\delta\left(1 - l_{\yb}\cdot P  \right)}{(l_{\xa}\cdot p_i)(l_{\yb}\cdot p_i)(l_{\xa}\cdot p_j)},
 \\
  & J_{20} && = \dInt {1,1,0}{1,1,0}{0}
  && = \int [\dm l_{\xa}] [\dm l_{\yb}] \frac{\delta\left(1 - l_{\xa}\cdot P\right)\delta\left(1 - l_{\yb}\cdot P  \right)}{(l_{\xa}\cdot p_i)(l_{\yb}\cdot p_i)(l_{\xa}\cdot p_j)(l_{\yb}\cdot p_j)},
 \\
  & J_{21} && = \dInt {0,0,1}{0,0,1}{0}
  && =\int [\dm l_{\xa}] [\dm l_{\yb}] \frac{\delta\left(1 - l_{\xa}\cdot P\right)\delta\left(1 - l_{\yb}\cdot P  \right)}{(l_{\xa\yb}\cdot p_i)(l_{\xa\yb}\cdot p_j)},
 \\
  & J_{22} && = \dInt {0,0,1}{0,0,2}{0}
  && = \int [\dm l_{\xa}] [\dm l_{\yb}] \frac{\delta\left(1 - l_{\xa}\cdot P\right)\delta\left(1 - l_{\yb}\cdot P  \right)}{(l_{\xa\yb}\cdot p_i)(l_{\xa\yb}\cdot p_j)^2},
 \\
  & J_{23} && = \dInt {0,0,1}{1,0,1}{0}
  && = \int [\dm l_{\xa}] [\dm l_{\yb}] \frac{\delta\left(1 - l_{\xa}\cdot P\right)\delta\left(1 - l_{\yb}\cdot P  \right)}{(l_{\xa\yb}\cdot p_i)(l_{\xa}\cdot p_j)(l_{\xa\yb}\cdot p_j)},
 \\
  & J_{24} && = \dInt {1,0,0}{0,0,1}{0}
  && = \int [\dm l_{\xa}] [\dm l_{\yb}] \frac{\delta\left(1 - l_{\xa}\cdot P\right)\delta\left(1 - l_{\yb}\cdot P  \right)}{(l_{\xa}\cdot p_i)(l_{\xa\yb}\cdot p_j)},
 \\
  & J_{25} && = \dInt {1,0,0}{0,0,2}{0}
  && = \int [\dm l_{\xa}] [\dm l_{\yb}] \frac{\delta\left(1 - l_{\xa}\cdot P\right)\delta\left(1 - l_{\yb}\cdot P  \right)}{(l_{\xa}\cdot p_i)(l_{\xa\yb}\cdot p_j)^2},
 \\
  & J_{26} && = \dInt {2,0,0}{0,0,1}{0}
  && = \int [\dm l_{\xa}] [\dm l_{\yb}] \frac{\delta\left(1 - l_{\xa}\cdot P\right)\delta\left(1 - l_{\yb}\cdot P  \right)}{(l_{\xa}\cdot p_i)^2(l_{\xa\yb}\cdot p_j)},
 \\
  & J_{27} && = \dInt {1,0,0}{0,0,1}{1}
  && = \int [\dm l_{\xa}] [\dm l_{\yb}] \frac{\delta\left(1 - l_{\xa}\cdot P\right)\delta\left(1 - l_{\yb}\cdot P  \right)}{(l_{\xa}\cdot p_i)(l_{\xa\yb}\cdot p_j)(l_{\xa}\cdot l_{\yb})},
 \\
  & J_{28} && = \dInt {1,0,0}{0,0,2}{1}
  && = \int [\dm l_{\xa}] [\dm l_{\yb}] \frac{\delta\left(1 - l_{\xa}\cdot P\right)\delta\left(1 - l_{\yb}\cdot P  \right)}{(l_{\xa}\cdot p_i)(l_{\xa\yb}\cdot p_j)^2(l_{\xa}\cdot l_{\yb})},
 \\
  & J_{29} && = \dInt {2,0,0}{0,0,1}{1}
  && = \int [\dm l_{\xa}] [\dm l_{\yb}] \frac{\delta\left(1 - l_{\xa}\cdot P\right)\delta\left(1 - l_{\yb}\cdot P  \right)}{(l_{\xa}\cdot p_i)^2(l_{\xa\yb}\cdot p_j)(l_{\xa}\cdot l_{\yb})},
 \\
  & J_{30} && = \dInt {1,0,1}{0,0,1}{0}
  && = \int [\dm l_{\xa}] [\dm l_{\yb}] \frac{\delta\left(1 - l_{\xa}\cdot P\right)\delta\left(1 - l_{\yb}\cdot P  \right)}{(l_{\xa}\cdot p_i)(l_{\xa\yb}\cdot p_i)(l_{\xa\yb}\cdot p_j)},
 \\
  & J_{31} && = \dInt {1,0,1}{0,0,2}{0}
  && = \int [\dm l_{\xa}] [\dm l_{\yb}] \frac{\delta\left(1 - l_{\xa}\cdot P\right)\delta\left(1 - l_{\yb}\cdot P  \right)}{(l_{\xa}\cdot p_i)(l_{\xa\yb}\cdot p_i)(l_{\xa\yb}\cdot p_j)^2},
 \\
  & J_{32} && = \dInt {1,0,2}{0,0,1}{0}
  && = \int [\dm l_{\xa}] [\dm l_{\yb}] \frac{\delta\left(1 - l_{\xa}\cdot P\right)\delta\left(1 - l_{\yb}\cdot P  \right)}{(l_{\xa}\cdot p_i)(l_{\xa\yb}\cdot p_i)^2(l_{\xa\yb}\cdot p_j)},
  \\
  & J_{33} && = \dInt {1,0,1}{1,0,1}{0}
  && = \int [\dm l_{\xa}] [\dm l_{\yb}] \frac{\delta\left(1 - l_{\xa}\cdot P\right)\delta\left(1 - l_{\yb}\cdot P  \right)}{(l_{\xa}\cdot p_i)(l_{\xa\yb}\cdot p_i)(l_{\xa}\cdot p_j)(l_{\xa\yb}\cdot p_j)},
  \\
  & J_{34} && = \dInt {1,1,0}{0,0,1}{0}
  && = \int [\dm l_{\xa}] [\dm l_{\yb}] \frac{\delta\left(1 - l_{\xa}\cdot P\right)\delta\left(1 - l_{\yb}\cdot P  \right)}{(l_{\xa}\cdot p_i)(l_{\yb}\cdot p_i)(l_{\xa\yb}\cdot p_j)},
  \\
  & J_{35} && = \dInt {1,1,0}{0,0,2}{0}
  && = \int [\dm l_{\xa}] [\dm l_{\yb}] \frac{\delta\left(1 - l_{\xa}\cdot P\right)\delta\left(1 - l_{\yb}\cdot P  \right)}{(l_{\xa}\cdot p_i)(l_{\yb}\cdot p_i)(l_{\xa\yb}\cdot p_j)^2},
  \\
  & J_{36} && = \dInt {1,2,0}{0,0,1}{0}
  && = \int [\dm l_{\xa}] [\dm l_{\yb}] \frac{\delta\left(1 - l_{\xa}\cdot P\right)\delta\left(1 - l_{\yb}\cdot P  \right)}{(l_{\xa}\cdot p_i)(l_{\yb}\cdot p_i)^2(l_{\xa\yb}\cdot p_j)},
  \\
  & J_{37} && = \dInt {1,1,0}{1,0,1}{0}
  && = \int [\dm l_{\xa}] [\dm l_{\yb}] \frac{\delta\left(1 - l_{\xa}\cdot P\right)\delta\left(1 - l_{\yb}\cdot P  \right)}{(l_{\xa}\cdot p_i)(l_{\yb}\cdot p_i)(l_{\xa}\cdot p_j)(l_{\xa\yb}\cdot p_j)},
  \\
  %
   %
  & J_{38} && = \tInt {0,0,1}{0,0,0}{0}
  && = \int [\dm l_{\xa}] [\dm l_{\yb}] \frac{\delta\left(1 - l_{\xa}\cdot P\right) \theta\left(l_{\xa}\cdot P - l_{\yb}\cdot P  \right)}{(l_{\xa\yb}\cdot p_i)},
  \\
  & J_{39} && = \tInt {0,0,2}{0,0,0}{0}
  && = \int [\dm l_{\xa}] [\dm l_{\yb}] \frac{\delta\left(1 - l_{\xa}\cdot P\right) \theta\left(l_{\xa}\cdot P - l_{\yb}\cdot P  \right)}{(l_{\xa\yb}\cdot p_i)^2},
  \\
  & J_{40} && = \tInt {0,0,1}{0,1,0}{1}
  && = \int [\dm l_{\xa}] [\dm l_{\yb}] \frac{\delta\left(1 - l_{\xa}\cdot P\right) \theta\left(l_{\xa}\cdot P - l_{\yb}\cdot P  \right)}{(l_{\xa\yb}\cdot p_i)(l_{\yb}\cdot p_j)(l_{\xa}\cdot l_{\yb})},
  \\
  & J_{41} && = \tInt {0,0,1}{0,2,0}{1}
  && = \int [\dm l_{\xa}] [\dm l_{\yb}] \frac{\delta\left(1 - l_{\xa}\cdot P\right) \theta\left(l_{\xa}\cdot P - l_{\yb}\cdot P  \right)}{(l_{\xa\yb}\cdot p_i)(l_{\yb}\cdot p_j)^2(l_{\xa}\cdot l_{\yb})},
  \\
  & J_{42} && = \tInt {0,0,2}{0,1,0}{1}
  && = \int [\dm l_{\xa}] [\dm l_{\yb}] \frac{\delta\left(1 - l_{\xa}\cdot P\right) \theta\left(l_{\xa}\cdot P - l_{\yb}\cdot P  \right)}{(l_{\xa\yb}\cdot p_i)^2(l_{\yb}\cdot p_j)(l_{\xa}\cdot l_{\yb})},
  \\
  & J_{43} && = \tInt {0,0,1}{0,0,1}{0}
  && = \int [\dm l_{\xa}] [\dm l_{\yb}] \frac{\delta\left(1 - l_{\xa}\cdot P\right) \theta\left(l_{\xa}\cdot P - l_{\yb}\cdot P  \right)}{(l_{\xa\yb}\cdot p_i)(l_{\xa\yb}\cdot p_j)},
  \\
  & J_{44} && = \tInt {0,0,1}{0,0,2}{0}
  && = \int [\dm l_{\xa}] [\dm l_{\yb}] \frac{\delta\left(1 - l_{\xa}\cdot P\right) \theta\left(l_{\xa}\cdot P - l_{\yb}\cdot P  \right)}{(l_{\xa\yb}\cdot p_i)(l_{\xa\yb}\cdot p_j)^2},
  \\
  & J_{45} && = \tInt {0,1,0}{0,0,1}{0}
  && = \int [\dm l_{\xa}] [\dm l_{\yb}] \frac{\delta\left(1 - l_{\xa}\cdot P\right) \theta\left(l_{\xa}\cdot P - l_{\yb}\cdot P  \right)}{(l_{\yb}\cdot p_i)(l_{\xa\yb}\cdot p_j)},
  \\
  & J_{46} && = \tInt {0,1,0}{0,0,1}{1}
  && = \int [\dm l_{\xa}] [\dm l_{\yb}] \frac{\delta\left(1 - l_{\xa}\cdot P\right) \theta\left(l_{\xa}\cdot P - l_{\yb}\cdot P  \right)}{(l_{\yb}\cdot p_i)(l_{\xa\yb}\cdot p_j)(l_{\xa}\cdot l_{\yb})},
  \\
  & J_{47} && = \tInt {0,1,0}{0,0,2}{1}
  && = \int [\dm l_{\xa}] [\dm l_{\yb}] \frac{\delta\left(1 - l_{\xa}\cdot P\right) \theta\left(l_{\xa}\cdot P - l_{\yb}\cdot P  \right)}{(l_{\yb}\cdot p_i)(l_{\xa\yb}\cdot p_j)^2(l_{\xa}\cdot l_{\yb})},
  \\
  & J_{48} && = \tInt {0,2,0}{0,0,1}{1}
  && = \int [\dm l_{\xa}] [\dm l_{\yb}] \frac{\delta\left(1 - l_{\xa}\cdot P\right) \theta\left(l_{\xa}\cdot P - l_{\yb}\cdot P  \right)}{(l_{\yb}\cdot p_i)^2(l_{\xa\yb}\cdot p_j)(l_{\xa}\cdot l_{\yb})},
  \\
  & J_{49} && = \tInt {1,0,1}{0,0,1}{0}
  && = \int [\dm l_{\xa}] [\dm l_{\yb}] \frac{\delta\left(1 - l_{\xa}\cdot P\right) \theta\left(l_{\xa}\cdot P - l_{\yb}\cdot P  \right)}{(l_{\xa}\cdot p_i)(l_{\xa\yb}\cdot p_i)(l_{\xa\yb}\cdot p_j)},
  \\
  & J_{50} && = \tInt {1,0,1}{0,0,2}{0}
  && =\int [\dm l_{\xa}] [\dm l_{\yb}] \frac{\delta\left(1 - l_{\xa}\cdot P\right) \theta\left(l_{\xa}\cdot P - l_{\yb}\cdot P  \right)}{(l_{\xa}\cdot p_i)(l_{\xa\yb}\cdot p_i)(l_{\xa\yb}\cdot p_j)^2},
  \\
  & J_{51} && = \tInt {1,1,0}{0,0,1}{0}
  && = \int [\dm l_{\xa}] [\dm l_{\yb}] \frac{\delta\left(1 - l_{\xa}\cdot P\right) \theta\left(l_{\xa}\cdot P - l_{\yb}\cdot P  \right)}{(l_{\xa}\cdot p_i)(l_{\yb}\cdot p_i)(l_{\xa\yb}\cdot p_j)},
  \\
  & J_{52} && = \tInt {1,1,0}{0,0,2}{0}
  && = \int [\dm l_{\xa}] [\dm l_{\yb}] \frac{\delta\left(1 - l_{\xa}\cdot P\right) \theta\left(l_{\xa}\cdot P - l_{\yb}\cdot P  \right)}{(l_{\xa}\cdot p_i)(l_{\yb}\cdot p_i)(l_{\xa\yb}\cdot p_j)^2}.
\end{alignat}

\section{Boundary conditions}
\label{sec:bcCalc}

Here we list integrals used to compute the boundary conditions at $\beta =0$. 

\begin{align}
  \label{eq:bcBdef}
J_{1} = & \Nep^2,\\
\frac{J_{2}}{J_{1}} = &
                  \frac{2^{-1-2 \ep} (1-2 \ep)^2 \Gamma^4 (1-2 \ep)
                  \Gamma (\ep)}{(1-4 \ep) \Gamma (1-4 \ep) \Gamma^3 (1-\ep)}
                  - \frac{(1-2 \ep)}{2 \ep}\, {}_3F_2(1,1-\ep,2 \ep;2-2 \ep,1 + \ep;-1)
  \\
 = & 
 \log (2)+\ep  \left(\frac{\pi ^2}{12}-\log ^2(2)\right)+\ep ^2 \left(-\zeta _3+\frac{2 \log ^3(2)}{3}-\frac{1}{6} \pi ^2 \log (2)+4 \log (2)\right) \nonumber \\
  & +\ep ^3 \left(-12 \zeta _3 \log (2)-16
   \text{Li}_4\left(\frac{1}{2}\right)+\frac{5 \pi ^4}{144}+\frac{\pi ^2}{3}-\log ^4(2) \right. \nonumber \\
  & \qquad\qquad \left.  +\frac{5}{6} \pi ^2 \log ^2(2)-4 \log ^2(2)+16 \log (2)\right)+\mathcal{O}\left(\ep ^4\right)
 , \nonumber \\
\frac{J_{38}}{J_{1}} =& \frac{1}{2}\left(\Psi\left(3/2 - \ep\right) - \Psi\left(1 - \ep\right)\right)  \\
 = & 
 (1-\log (2))+ \ep\left(2-\frac{\pi ^2}{6}\right) +\ep ^2\left(4-3 \zeta _3\right) +\ep ^3\left(8-\frac{7 \pi ^4}{90}\right) +\mathcal{O}\left(\ep ^4\right)
 , \nonumber \\
\frac{J_{43}}{J_{1}} =& \int\limits_0^1 \frac{\dm z}{1+z}\Biggl[
                  - z^{-2 \ep} \frac{(1-2 \ep)  }{2 \ep} \, {}_3F_2\left(1,1-\ep,2 \ep;2-2 \ep,1+\ep;-\frac{1}{z}\right)\\
                & + z^{- \ep} \frac{(1-2 \ep)^2  \Gamma (1-2 \ep)^3 \Gamma (\ep)}{2 (1 - 3 \ep) \Gamma (1-3 \ep) \Gamma^2 (1-\ep)} \, {}_2F_1\left(1-2 \ep,\ep;2-3\ep;-\frac{1}{z}\right) \Biggr]
                \nonumber 
                \\
                =& \log(2)-\frac{1}{24}\pi ^2
                  +\ep \left(
                  4 \log(2)
                  -\log^2(2)
                  +\frac{1}{12} \pi ^2
                  -\frac{17}{8} \zeta_3
                  \right)
                  \nonumber \\
                & +\ep^2 \left(
                  +20 \log(2)
                  -4 \log^2(2)
                  +\frac{1}{6} \pi^2
                  +\frac{2}{3} \log^3(2)
                  -\frac{1}{6} \pi^2 \log(2)
                  -\zeta_3\right.\nonumber\\
                  & \left.-\frac{5}{12} \log^4(2)
                  +\frac{1}{480} \pi^4
                  +\frac{5}{12} \pi^2 \log^2(2)
                  -\frac{35}{4} \zeta_3 \log(2)
                  -10 \textrm{Li}_4\left(\frac{1}{2}\right)
                  \right)
                  + \mathcal{O}\left(\ep^3\right),
 \nonumber  \\
 \frac{J_{45}}{J_{1}} = &
                  - \frac{1}{2 \ep}\, {}_4F_3(1,1-2 \ep,1-\ep,2 \ep;2-2 \ep,2-2 \ep,1+\ep;-1)\\
                &   + \frac{(1 - 2 \ep)^2 \Gamma^3 (1-2 \ep) \Gamma (\ep)}{2 (1 - 3 \ep)^2
                  \Gamma (1-3 \ep) \Gamma^2 (1-\ep)} \, {}_3F_2(1-3 \ep,1-2 \ep,\ep;2-3 \ep,2-3 \ep;-1)
                  \nonumber \\
 = & 
 \left(\frac{\pi ^2}{24}+\log (2)\right)+\ep  \left(\frac{5 \zeta _3}{8}+\frac{\pi ^2}{12}-\log ^2(2)+4 \log (2)\right) \nonumber  \\
 & +\ep ^2 \left(-\zeta _3-\frac{21}{4} \zeta _3 \log (2)-6 \text{Li}_4\left(\frac{1}{2}\right)+\frac{61
   \pi ^4}{1440}+\frac{\pi ^2}{2}-\frac{\log ^4(2)}{4}+\frac{2 \log ^3(2)}{3} \right. \nonumber  \\
   & \qquad \left. +\frac{1}{4} \pi ^2 \log ^2(2)-4 \log ^2(2)-\frac{1}{6} \pi ^2 \log (2)+20 \log (2)\right)+\mathcal{O}\left(\ep ^3\right).
   \nonumber 
\end{align}

\section{Angular integrals}
\label{sec:angIntsDef}

In this appendix, we define angular integrals which are used for calculations in
Section~\ref{sec:calcRestA}.

\begin{alignat}{3}
  \renewcommand\arraystretch{1.33}
  \label{eq:angIntsBrackets}
  & \Big \langle 1 \Big \rangle_{\xa} = \Big \langle 1 \Big \rangle_{\xa \yb} && = 1,&&\\
  & \Big \langle \frac{1}{\rho_{\xa x}^n} \Big \rangle_{\xa} && = \AIm{n}\left[\rho_{xx}\right], &&\rho_{xx} \neq 0,\\
  & \Big \langle \frac{1}{\rho_{\xa x}^a \rho_{\xa y}^b} \Big \rangle_{\xa} && = \AIzz{a}{b}{\rho_{xy}} , &&\rho_{xx} = \rho_{yy} = 0,\\
  & \Big \langle \frac{1}{\rho_{\xa x}^a \rho_{\xa y}^b} \Big \rangle_{\xa} && = \AImz{a}{b}\left[ \rho_{xx}, \rho_{xy} \right] ,\quad&& \rho_{xx} \neq 0, \rho_{yy} = 0.
\end{alignat}
These integrals can be computed in terms of hypergeometric functions and the Appell function
\begin{align}
  &\AIm{n}\left[ \rho_{11} \right]
   = \left( 1 + \sqrt{1-\rho_{11}} \right)^{-n}  {}_2F_1\left(n,1-\ep,2-2\ep; \frac{2 \sqrt{1-\rho_{11}}}{1+\sqrt{1-\rho_{11}}}  \right) \label{eq:angInt-1-m},\\
  &\AIzz{a}{b}{\rho_{12}}
   = \frac{\Gamma\left(2-2\ep\right)\Gamma\left(1-\ep -a \right)\Gamma\left(1-\ep -b \right)}{2^{a+b}\Gamma^2\left(1-\ep\right)\Gamma\left(2-2\ep - a -b \right)}
    {}_2F_1\left(a,b, 1-\ep; 1-\frac{\rho_{12}}{2} \right), \label{eq:angInt-2-00}\\
 &\AImz{a}{b}\left[ \rho_{11}, \rho_{12} \right]
   = \frac{2^{-b}}{\rho_{12}^{a}} \frac{\Gamma\left(2-2\ep\right)\Gamma\left(1-\ep-b \right)}{\Gamma\left(1-\ep \right)\Gamma\left(2-2\ep-b  \right)} \times 
\label{eq:angInt-2-0m}
   \\
  & \quad F_1\left(
    a, 1-\ep-b, 1-\ep-b,2-2\ep-b
    ;1-\frac{1+\sqrt{1-\rho_{11}}}{\rho_{12}}, 1-\frac{1-\sqrt{1-\rho_{11}}}{\rho_{12}} \right).
    \nonumber 
\end{align}
Expansions of all integrals needed for computations in
Section~\ref{sec:calcRestA} in $\ep$ to the required order can be found in
ancillary files. Here, we present a few terms of such an expansion for
illustration purposes. To present the result, we use variables $x=\beta$ and $y
= \cos \theta$ that refer to the $P = (1,\vec 0)$ frame. In the massive-parton
rest frame, we find
\begin{equation}
  \label{eq:rho2xy}
  \rho_{tt} = 1 - x^2 = 1 - \beta^2, \quad \rho_{tj} = \frac{1-\beta^2}{1-\beta \csth} = \frac{1-x^2}{1-x y}. 
\end{equation}
The results for integrals read
\begin{align}
  \label{eq:AIepExp}
  &\AIm{\alpha\ep}\left[ \rho_{tt} \right]
  =
    1 +
    \alpha \ep \left(1  + \frac{1-x}{2x} \log (1-x) -\frac{1+x}{2x} \log (1+x) \right) \\
  & +\ep^2
    \Biggl(
    \alpha(\alpha+2)   -\frac{\pi ^2 \alpha}{6 x}
    +\frac{1}{2} \alpha^2 \log ^2(1-x)
    -\alpha^2 \log (1-x)
    \nonumber \\
  &+ \frac{\alpha}{x} \left(
    \frac{ (\alpha (1+x) + 2\log(2))}{2}
    -\frac{(\alpha (1+x)+2) \log (1-x)}{2}
    +\log (x)\right) \log \left(\frac{1-x}{1+x}\right) \nonumber    
  \\
  & +\frac{\alpha (\alpha (1+x) + x+3)}{4 x} \log^2\left(\frac{1-x}{1+x}\right)
     +\frac{\alpha}{x} \mathrm{Li}_2\left(\frac{1-x}{1+x}\right)
    \Biggr) + \mathcal{O}\left( \ep^3 \right),
 \nonumber  \\
  & \AImz{\alpha \ep}{1}\left[ \rho_{tt}, \rho_{tj} \right] =
    -\frac{1}{2\ep} + 1 +\frac{\alpha}{2} \log{\frac{1-x^2}{1-x y}}
    + \ep \Biggl[
    \alpha \biggl(
    -\frac{1}{2} \log^2\left(\frac{1-x^2}{1 - x y}\right)\\
  &+\frac{1}{2} \log (1-x^2) \log \left(\frac{1 - x^2}{1 - x y}\right)
    - \log \left(\frac{1 - x^2}{1 - x y}\right)
    +\frac{1}{2} \log \left(\frac{1-x}{1-x y}\right) \log \left(\frac{x (1-y)}{1 - x y}\right)
    \nonumber \\
    & +\frac{1}{2} \log \left(\frac{x+1}{1-x y}\right) \log \left(\frac{x (y+1)}{x y-1}\right)
      -\frac{1}{4} \log ^2(1-x)-3 \log ^2(x+1)-2 \pi^2
      \nonumber 
      \\
  &
    + \frac{1}{2} \mathrm{Li}_2\left(\frac{1-x}{1-x y}\right)
    +\frac{1}{2} \mathrm{Li}_2\left(\frac{1+x}{1-x y}\right)
    \biggr)
    -\frac{1}{4} \alpha^2 \log ^2\left(\frac{1-x^2}{1 - x y}\right)\Biggr] + \mathcal{O}\left( \ep^2 \right),
 \nonumber  \\
  & \AImz{\alpha \ep}{\ep}\left[ \rho_{tt}, \rho_{tj} \right] =
    1+\ep \Biggl[
    1-\log(2) +
    \alpha\biggl(
    1
    +\frac{1}{2x}\left(
    +\log \frac{1-x}{1+x} -x \log (1-x^2)
    \right)
    \biggr)
    \Biggr] \\
  &
    +\ep^2\Biggl[
    -\frac{\pi ^2}{6}+3+\frac{1}{2} (\log (2)-2) \log (2)
    +\alpha \Biggl(
    -\frac{1}{2x}\biggl(
    -8 x
    -\left((x-1) \log ^2(1-x)\right)
    \nonumber \\
  &
    -(x+1) \log ^2(x+1)
    -2 x \log ^2(2)
    +(x-1) (1+\log (2)) \log (1-x)
   \nonumber  \\
  &
    +(x \log (2)+1+\log (2)) \log(x+1)
    +x \log (4 (x+1))
    \nonumber \\
  &    
    + (1+x) \log \left(\frac{x-1}{x y-1}\right) \log \left(\frac{x (y-1)}{x y-1}\right)
    +(x-1) \log \left(\frac{x+1}{1-x y}\right) \log \left(\frac{x (y+1)}{x y-1}\right)
    \biggr)
\nonumber \\
  &    -\frac{(1+x)}{2 x}\mathrm{Li}_2\left(\frac{1-x}{1-x y}\right)
    +\frac{(1-x)}{2x}  \mathrm{Li}_2\left(\frac{1+x}{1 - x y}\right)
        +\frac{2}{x}\mathrm{Li}_2(-x)
    -\frac{2}{x}\mathrm{Li}_2(x)
    \nonumber \\
    & -\frac{(1-x)}{x} \mathrm{Li}_2\left(\frac{1-x}{2}\right)
    +\frac{(1+x)}{x} \mathrm{Li}_2\left(\frac{1+x}{2}\right)
    \Biggr)
      + \frac{\alpha^2}{4 x} \biggl(4 x+(x-1) \log ^2(1-x)
      \nonumber \\
  &
      -2 (x-1) \log (1-x) +(x+1) (\log (x+1)-2) \log (x+1)\biggr)
    \Biggr]
    + \mathcal{O}\left(\ep^3\right),
 \nonumber  \\
  & \AImz{\alpha \ep}{1+\ep}\left[ \rho_{tt}, \rho_{tj} \right]=
    -\frac{1}{4 \ep}
    +\frac{1}{4} \left(
    2+\log (2) + \alpha \log \left(\frac{1-x^2}{1-x y}\right)
    \right)\\
  & + \ep \Biggl[
    \frac{1}{24} \left(\pi ^2-3 \log (2)(4+\log (2))\right)
    +\alpha \biggl(
    \frac{1}{2} \mathrm{Li}_2\left(\frac{1-x}{1 - x y}\right)
    +\frac{1}{2}\mathrm{Li}_2\left(\frac{1 + x}{1 - x y}\right)
    \nonumber \\
  & 
    -\frac{1}{2} \log ^2\left(\frac{1 - x^2}{1 - x y}\right)
    +\frac{1}{2} \log (1-x^2) \log \left(\frac{1 - x^2}{1 - x y}\right)
    -\frac{(2+ \log (2))}{4} \log \left(\frac{1 - x^2}{1 - x y}\right)
    \nonumber \\
  & 
    +\frac{1}{2} \log \left(\frac{1 - x}{1 - x y}\right) \log \left(\frac{x (1-y)}{1 - x y}\right)
    +\frac{1}{2} \log \left(\frac{1 + x}{1-x y}\right) \log \left(\frac{x (1 + y)}{x y-1}\right)
    \nonumber \\
  & 
    -\frac{1}{4} \log^2(1-x)
    -\frac{1}{4} \log^2(x+1)
    -\frac{1}{6} \pi^2
    \biggr)
   -\frac{1}{8} \alpha^2 \log ^2\left(\frac{x^2-1}{x y-1}\right) 
    \Biggr]
    + \mathcal{O}\left(\ep^2\right).\nonumber
\end{align}

\section{Direct integration}
\label{sec:directInt}

To check master integrals, we compute them numerically in higher-dimensional
space-time where they are finite. We use the following parameterization of
vectors
\renewcommand\arraystretch{1.33} 
\begin{equation}
\scalemath{0.98}{
  \begin{array}{l @{{}={}} c @{{},{}} c @{{},{}} c @{{},{}} c @{{},{}} c @{{},{}}  c @{{}{}} c}
    p_i & \phantom{z}\bigl( 1 &\ \vec{0}_{d-5} & 0 & 0  & \beta\sin{\ijangle} & \beta \cos{\ijangle} &\bigr),\\
    p_j & \phantom{z}\bigl( 1 &\ \vec{0}_{d-5} & 0 & 0  & 0              & 1 &\bigr),\\
    l_1 & \phantom{z}\bigl( 1 &\ \vec{0}_{d-5} & 0 & \ \sin{\theta_{11}}\sin{\theta_{12}} & \ \sin{\theta_{11}}\cos{\theta_{12}} & \ \cos{\theta_{11}} & \bigr),\\
    l_2 & z \bigl(1  &\ \vec{0}_{d-5} & \ \sin{\theta_{21}} \sin{\theta_{22}}\sin{\theta_{23}}   & \ \sin{\theta_{21}}\sin{\theta_{22}}\cos{\theta_{23}} & \ \sin{\theta_{21}}\cos{\theta_{22}} & \ \cos{\theta_{21}} & \bigr).
  \end{array}
  }
\end{equation}

The integration measure reads
\begin{equation}
  \label{eq:dDk}
  \int \dm^d l \ \delta(l^2)\theta(E_l) f(l)= \int \frac{\dm^{d-1} \vec{l}}{2 E_l} f(l) = \int \frac{\dm E_l}{2 E_l^{3-d}} \int \dm \Omega^{(d-1)}_l f(l) .
\end{equation}
Parameterizing the relevant angles as 
 $\cos{\theta_{ij}} =
1-2\lambda_{ij},\;\; \sin{\theta_{ij}} = 2\sqrt{\lambda_{ij} \bar{\lambda}_{ij}}
$,
we write integration over solid angles as follows 
\begin{equation}
  \begin{split}
    \label{eq:dOmk1k2}
    &  \int \dm \Omega^{(d-1)}_1 \dm \Omega^{(d-1)}_2  = 32 \Omega^{(d-4)} \Omega^{(d-3)} \int\limits_0^1 \dm \lambda_{11} \dm \lambda_{12}\dm \lambda_{21}\dm \lambda_{22}\dm \lambda_{23} 
    \\
    & \times \Lambda_{11}^{d_0-4}\Lambda_{12}^{d_0-5}\Lambda_{21}^{d_0-4}\Lambda_{22}^{d_0-5}\Lambda_{23}^{d_0-6} \sum\limits_{k=0}^\infty \ep^k \frac{\left[ -\log{(\Lambda_{11}^{2}\Lambda_{12}^{2}\Lambda_{21}^{2}\Lambda_{22}^{2}\Lambda_{23}^{2})} \right]^k}{k!},
  \end{split}
\end{equation}
where $\Lambda_{ij} =2\sqrt{\lambda_{ij}(1-\lambda_{ij})}$ and $d = d_0 -2\ep$.
\section{Ancillary files}
The following ancillary files, in \texttt{Mathematica}-readable format,  are provided with this paper:
\begin{itemize}
    \item \texttt{I11} contains the expansion of the single-emission soft integral,
    defined in 
     Eq. (\ref{eq: mi_nlo_11}),
    up to ${\cal O}(\ep^3)$ in terms of conventional polylogarithms and $\textrm{Li}_{2,2}$; 
    \item \texttt{Im0exp} contains expansion of angle integrals $\AImz{a}{b}\left[\rho_{xx},\rho_{xy}\right]$ to higher orders
    in $\ep$, see Appendix \ref{sec:angIntsDef};
    \item \texttt{SSm0\_tldI} contains the integrated double-emission eikonal function $\intSS\left[ \widetilde{\mathcal{I}}_{ij} \right]$, with
    the normalization factor $-\Nep^2/(4\ep\emax^{4\ep})$ omitted, in terms of polylogarithms and $\textrm{Li}_{2,2}$;
    \item \texttt{SSm0\_tldS} contains the integrated double-emission eikonal function $\intSS\left[ \widetilde{\mathcal{S}}_{ij} \right]$, with 
    the normalization factor  $-\Nep^2/(4\ep\emax^{4\ep})$ omitted, in terms of polylogarithms and $\textrm{Li}_{2,2}$.
\end{itemize}
To retrieve the results for integrated eikonal functions, files
\texttt{SSm0\_tldI} and \texttt{SSm0\_tldS} have to loaded into a
\texttt{Mathematica} session, together with the package
\texttt{PolylogTools}\cite{Duhr:2019tlz}, since it provides an interface to
\texttt{GiNaC}\cite{Vollinga:2004sn} which is needed for ${\rm Li}_{2,2}$
evaluation. An example of a \texttt{Mathematica} session is shown below.
\begin{lstlisting}
(* Numerical values specify beta (xNum) and cos(theta) (yNum) values *)
With[{xNum = 0.2, yNum = 0.3},
(* Gluons, Quarks *)
   { Get["SSm0_tldS"],Get["SSm0_tldI"]}
    /. {eta -> (1 - 3*x*y - Sqrt[1 + 8*x^2 - 6*x*y + x^2*y^2])/x}
    /. {Li22[a_, b_] :> Ginsh[Li[{2, 2}, {b, a}], {x -> xNum, y -> yNum}]}
    /. {x -> xNum, y -> yNum}]
(* Produced output:
{
-(1/(4 ep^3)) + 0.00020226/ep^2 - 0.440986/ep + 0.0147285 + 2.08178 ep + O[ep]^2,
-(1/(12 ep^2)) + 0.102352/ep + 0.018446 + 0.0847638 ep + O[ep]^2
} *)
\end{lstlisting}
Here the output refers to $\intSS\left[ \widetilde{\mathcal{S}}_{ij} \right]$
and $\intSS\left[ \widetilde{\mathcal{I}}_{ij} \right]$, divided by the
normalization factor $-\Nep^2/(4\ep\emax^{4\ep})$, computed at the kinematic
point $\beta = 0.2$ and $\cos \theta = 0.3$.

We also provide an implementation of $\intSS\left[ \widetilde{\mathcal{S}}_{ij}
\right]$ and $\intSS\left[ \widetilde{\mathcal{I}}_{ij} \right]$ in a \texttt{C}
code. The code is available from the \texttt{Github} repository, and can be
obtained using the following command
\begin{center}
\texttt{git clone \href{https://github.com/apik/SSm0.git}{https://github.com/apik/SSm0.git}}
\end{center}
Our implementation supports fast and accurate numerical evaluation of the
relevant $\textrm{Li}_{2,3,4}(x)$ and $\textrm{Li}_{2,2}(x,y)$ functions with
machine precision. We use algorithms described in Ref.~\cite{Frellesvig:2016ske} with
modifications suitable for our case, where only real-valued functions are
involved and their arguments satisfy certain constraints. After downloading the
code from GitHub, one can build the library, create an executable, and run it
for the same point $\beta=0.2, \csth=0.3$ using the following commands
\begin{lstlisting}[language=bash]
$ make
$ ./ex_SSm0 0.2 0.3
\end{lstlisting}
The following output should then appear on the screen  
\begin{lstlisting}[language={}]
    beta       =   0.20000000
    cos(theta) =   0.30000000
                           \tilde{I}(quarks)                 \tilde{S}(gluons)
    ep^-3                      0.0000000000                     -0.2500000000
    ep^-2                     -0.0833333333                      0.0002022602
    ep^-1                      0.1023516932                     -0.4409857886
    ep^ 0                      0.0184460303                      0.0147284599
    ep^ 1                      0.0847637908                      2.0817758455
\end{lstlisting}
To obtain $\intSS\left[ \widetilde{\mathcal{S}}_{ij} \right]$ and $\intSS\left[
\widetilde{\mathcal{I}}_{ij} \right]$, one has to multiply these numbers by the
normalization factor $-\Nep^2/(4\ep\emax^{4\ep})$.

\bibliographystyle{JHEP}
\bibliography{SSm0}

\end{document}